# Image enhancement in acoustic-resolution photoacoustic microscopy enabled by a novel directional algorithm

Fei Feng[a+], Siqi Liang[a+], Sung-Liang Chen[a,b,c*]

[a]University of Michigan-Shanghai Jiao Tong University Joint Institute, Shanghai Jiao Tong University, Shanghai 200240, China
[b]Engineering Research Center of Digital Medicine and Clinical Translation, Ministry of Education, Shanghai 200030, China
[c]State Key Laboratory of Advanced Optical Communication Systems and Networks, Shanghai Jiao Tong University, Shanghai 200240, China

[+]: These authors contributed equally to this work.
*: Corresponding author. E-mail address: sungliang.chen@sjtu.edu.cn (S.-L. Chen)

**Abstract**: Acoustic-resolution photoacoustic microscopy (AR-PAM) is a promising tool for microvascular imaging. In the focal region, resolution of AR-PAM is determined by the ultrasound transducer and ultimately limited by acoustic diffraction. In the out-of-focus region, resolution deteriorates with increasing distance from the focal plane, which restricts depth of focus (DOF). Besides, a trade-off exists between resolution and DOF. Previously, synthetic aperture focusing technique (SAFT) and/or deconvolution methods have been demonstrated to enhance AR-PAM images. However, they suffer from issues in low resolution, low signal-to-noise ratio (SNR), and/or poor image fidelity. Here, we propose a novel algorithm for AR-PAM to enhance image resolution, SNR, and fidelity. The algorithm consists of a Fourier accumulation SAFT (FA-SAFT) and a directional model-based (D-MB) deconvolution method. Inspired from Fourier denoising technique and directional SAFT, FA-SAFT mainly compensates for the defocusing effect. Besides, D-MB deconvolution enhances the resolution as well as preserves the image fidelity, especially for the objects with line patterns such as microvasculature. Full width at half maximum of 26−31 μm over DOF of 1.8 mm and minimum resolvable distance of 46−49 μm are experimentally achieved by imaging tungsten wire phantom. Moreover, imaging of leaf skeleton phantom and *in vivo* imaging of mouse blood vessels also prove that our algorithm is capable of providing high-resolution, high-SNR, and good-fidelity results for complex structures and for *in vivo* applications.

## 1. Introduction

Photoacoustic (PA) imaging (PAI) has developed rapidly in the past two decades. In PA image acquisition, an object is excited by a light source, and ultrasound (US) waves emerge because of the PA effect. Then, one can recover the light absorber distribution by collecting the US waves with an US transducer and using a reconstruction algorithm. PAI has achieved promising results in microvascular imaging [1−3] and endoscopic imaging [4], to name a few. PA microscopy (PAM) is one implementation of PAI and can be classified into optical-resolution PAM (OR-PAM) and acoustic-resolution PAM (AR-PAM), depending on the focusing approach. Overall, OR-PAM enjoys high resolution but suffers from low imaging depth (~1 mm) while AR-PAM can achieve deep imaging depth by taking advantage of diffused light and good acoustic penetration [5].

In AR-PAM, its lateral resolution depends on the center frequency and numerical aperture (NA) of a focused US transducer. High lateral resolution requires that the transducer has high center frequency and high NA, which in turn limit penetration and depth of focus (DOF) of AR-PAM. The former is because high-frequency US waves attenuate dramatically in biological tissue, and the latter is because a high-NA transducer has limited DOF. Besides, the ultimate resolution of AR-PAM is restricted by acoustic diffraction. Fortunately, previous works have demonstrated image enhancement in lateral resolution, DOF, and signal-to-noise ratio (SNR) by using algorithms, which circumvents the issues associated with the high-frequency and high-NA transducer.

Synthetic aperture focusing technique (SAFT) is often used in AR-PAM reconstruction for enhancing lateral resolution in the out-of-focus region. SAFT was first developed for US imaging to realize beamforming [6] and then was applied to PAI based on a point-like acoustic detector [7]. Later, Li *et al.* proposed the virtual-point-detector (VPD) concept to apply SAFT to AR-PAM using a focused transducer [8]. Afterward, SAFT was extended to two dimensions by Deng *et al.* [9]. Considering the line pattern of blood vessels in AR-PAM images, adaptive SAFT (A-

SAFT) was further investigated [10]. To improve resolution in the focal region, spatial-dependent contributions were weighted using the spatial impulse response (SIR) in SAFT [11]. More recently, delay-multiply-and-sum-based SAFT [12], two-dimensional (2D) spherical SAFT (S-SAFT) [13], and directional SAFT (D-SAFT) [14] have been further studied to improve lateral resolution and SNR in the out-of-focus region. Lateral resolution in the out-of-focus region can be effectively improved by SAFT but is still ultimately limited to the in-focus lateral resolution. Therefore, deconvolution-based algorithms are used to further improve lateral resolution in PAI [13,15,16] and US imaging [17,18]. Specifically, in AR-PAM, the point spread function (PSF) is dominated by the finite US detection beam size of a transducer. Deconvolution of acquired PA images with PSF could recover the raw PA distribution. Previously, Cai *et al.* [13,16] used Richardson-Lucy (R-L) deconvolution algorithm to improve the AR-PAM lateral and axial resolutions in both the focal and out-of-focus regions. However, the nature of the line pattern of blood vessels was not considered, and thus, the recovered lines suffer the discontinuity issue. An alternative model-based (MB) deconvolution method assumes the acquired PA image is the result of convolution of a PA source (an absorber) and a dictionary matrix. Then, the PA source distribution could be recovered by using an optimization approach. It has been demonstrated that the MB deconvolution method can improve resolution in both PAI and US imaging [19–21]. Besides, prior knowledge of sparsity was considered as a regularization term in the model to further enhance the resolution [20,21]. As a matter of fact, the nature of the line pattern of blood vessels in AR-PAM images has only one-dimensional (1D) sparsity (in contrast to 2D sparsity in [20,21]). If it is not considered, results will have the discontinuity issue, as also encountered in the R-L deconvolution algorithm mentioned above.

To overcome the above challenges, we propose a novel two-step algorithm including a Fourier accumulation SAFT (FA-SAFT) and a directional MB (D-MB) deconvolution method to improve resolution, SNR, and image fidelity in both the focal and out-of-focus regions for AR-PAM [22]. Our contributions can be summarized in two aspects. (i) FA-SAFT uses a sharpening filter to improve the high-frequency components of PA images in Fourier domain (k-space). Such a sharpening filter is adaptive to different patterns of the acquired PA images, leading to enhanced resolution and SNR. (ii) D-MB deconvolution takes 1D sparsity of the line pattern into account, which further enhances resolution and keeps image fidelity. The results by our algorithm are promising. By imaging tungsten wires, high resolution over large DOF and good image fidelity are experimentally achieved using our algorithm. Further, *in vivo* imaging is also demonstrated to show the superior capability of our algorithm. Compared with existing algorithms (D-SAFT, R-L deconvolution, etc.), our algorithm enables great imaging performance in resolution, SNR, and fidelity for both phantom and *in vivo* imaging demonstrations. The effect of SNR on the performance of our algorithm is also investigated and discussed.

**2. Related Techniques**
We first describe some existing techniques related to our algorithm. The conventional SAFT algorithms are introduced in Sections 2.1. Then, adaptive-based SAFT algorithms including A-SAFT and D-SAFT are introduced in Sections 2.2. Finally, R-L deconvolution and MB deconvolution are introduced in Sections 2.3.

*2.1. Conventional SAFT*
In AR-PAM, linear scanning is conducted in one dimension or two dimensions. We first illustrate SAFT for the 1D case, as shown in Fig. 1. In SAFT, signal synthesis is performed based on a delay-and-sum algorithm along the scanning line, which forms a synthetic aperture, and can be expressed as:

$$\text{RF}_{\text{SAFT}}(t) = \sum_{i=0}^{N-1} \text{RF}_i(t - \Delta t_i), \tag{1}$$

where $\text{RF}_{\text{SAFT}}(t)$ is the synthesized PA A-line signal after SAFT at a certain scanning position, $\text{RF}_i(t)$ is the signal acquired at the $i^{\text{th}}$ scanning position, and $\Delta t_i$ is the corresponding time delay between the target $\text{RF}_{\text{SAFT}}(t)$ at that certain scanning position and $\text{RF}_i(t)$. That is, when $\Delta t_i = 0$, the target $\text{RF}_{\text{SAFT}}(t)$ signal and the $\text{RF}_i(t)$ signal are at the same scanning position. Since AR-PAM is usually implemented with a focused transducer (Fig. 1b), a VPD concept was proposed, and the time delay is computed using the following equation [6], [11], [14]:

$$\Delta t_i = sgn(z - z_f) * \frac{r - r'}{c}, \tag{2}$$

where *sgn* is the sign function, and $z$ and $z_f$ are the axial distance of $\text{RF}_i(t)$ and the focal length of AR-PAM system, respectively. $z - z_f$ is the out-of-focus distance (OFD). $r$ and $r'$ are the distance to the focal point from the target $\text{RF}_{\text{SAFT}}(t)$ and from $\text{RF}_i(t)$, respectively. $c$ is the speed of sound in a propagation medium.

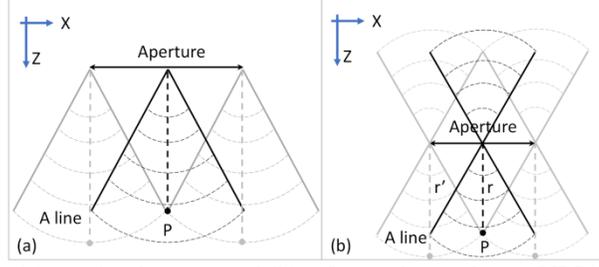

Fig. 1. Schematic of SAFT for (a) a point detector and (b) a focused transducer (VPD) in one dimension. Scanning is conducted along the X direction.

Due to the spatial response of the focused transducer, the previous study further proposed improved SAFT algorithm by incorporating SIR to weigh the spatial-dependent contributions of $RF_i(t)$ [11]. Eq. (1) is thus modified as:

$$RF_{SAFT}(t) = \sum_{i=0}^{N-1} RF_i(t - \Delta t_i) * SIR_i(t - \Delta t_i).$$

(3)

Furthermore, to suppress the out-of-phase summation in Eq. (1), coherence factor (CF) [7] is applied in SAFT and is defined as:

$$CF(t) = \frac{\left|\sum_{i=0}^{N-1} RF_i(t - \Delta t_i) * SIR_i(t - \Delta t_i)\right|^2}{N \sum_{i=0}^{N-1} (RF_i(t - \Delta t_i) * SIR_i(t - \Delta t_i))^2}$$

(4)

$$RF_{SAFT-CF}(t) = RF_{SAFT}(t) * CF(t).$$

(5)

1D SAFT can be extended to 2D SAFT by conducting the scanning in two dimensions. 2D cross SAFT (C-SAFT) [9] and 2D S-SAFT [13] were proposed to extend the effective aperture size, which is discussed in detail in previous literature [9,13].

*2.2. A-SAFT and D-SAFT*

AR-PAM is mostly used for vascular imaging. Considering the directional characteristic of the line pattern of blood vessels, A-SAFT and D-SAFT were proposed by Deng *et al.* [10] and Jeon *et al.* [14], respectively, aiming at SAFT for vascular images. A-SAFT is based on 1D SAFT with the signal synthesis directions adaptively changed with the directions of vessel line branches. However, the performance of A-SAFT depends on the successful recognition of the vessel branches' directions before applying A-SAFT. In other words, if one branch is not observed or recovered properly before A-SAFT is applied, it will not be recovered properly after A-SAFT.

To deal with this problem, D-SAFT was designed to recover vessel line branches in a directional way [14]. In D-SAFT, 1D SAFT along a series of directions is first applied, as described in the following equation:

$$CF(t) = \frac{\left|\sum_{i=0}^{N-1} RF_{i,\theta}(t - \Delta t_i) * SIR_{i,\theta}(t - \Delta t_i)\right|^2}{N \sum_{i=0}^{N-1} (RF_{i,\theta}(t - \Delta t_i) * SIR_{i,\theta}(t - \Delta t_i))^2}$$

(6)

$$RF_{SAFT-CF,\theta}(t) = \sum_{i=0}^{N-1} RF_{i,\theta}(t - \Delta t_i) * SIR_i(t - \Delta t_i) * CF(t),$$

(7)

where $\theta$ indicates the synthesis directions, which lies in the lateral plane and means the direction along which the 1D SAFT is conducted. For a specific $\theta$, vessel branches perpendicular to the $\theta$ direction can be well recovered. Therefore, different branches with diverse orientations can be recovered with different $\theta$. Then, D-SAFT uses the hamming window in k-space to compose the results from all directions. Three-dimensional (3D) D-SAFT procedure can be described using the equations:

$$K_\theta(u, v, \omega) = F\{RF_{SAFT-CF,\theta}(x, y, t)\}$$

(8)

$$K_{merged}(u,v,\omega) = \sum_{n=0}^{N'-1} K_{\theta_n}(u,v,\omega) * W_{\theta_n}(u,v,\omega),$$

(9)

where $F$ represents the Fourier transform, $K_\theta$ represents the spectrum of $RF_{SAFT-CF,\theta}$, and $u,v$ represents wavenumbers, corresponding to the coordinates $x,y$ in the lateral plane, respectively. $\omega$ represents PA signal frequency. $N'$ represents the number of directions for 1D SAFTs. $W_{\theta_n}$ is the weight for a specific direction $\theta_n$, and $K_{merged}$ is the combined results in k-space. As discussed above, for a certain branch direction, only the synthesis direction nearly perpendicular to the branch direction can properly recover the branch. $W_{\theta_n}(u,v,\omega)$ is used to weight the contribution of signal synthesis in different directions over the lateral plane in k-space ($K_\theta(u,v,\omega)$) and thus has no change over different $\omega$. Therefore, $W_{\theta_n}(u,v,\omega)$ degenerates to $W_{\theta_n}(u,v)$, which is expressed as follows [14]:

$$W_{\theta_n}(u,v) = \begin{cases} \dfrac{1}{N'}, & \text{if } u = \left\lceil \dfrac{X-1}{2} \right\rceil \text{ and } v = \left\lceil \dfrac{Y-1}{2} \right\rceil \\ \cos^2\left( \dfrac{\overline{\theta_n}(u,v) * N'}{2} \right), & \text{if } |\overline{\theta_n}(u,v)| \leq \dfrac{\pi}{N'} \\ 0, & \text{otherwise} \end{cases}$$

(10)

$$\overline{\theta_n}(u,v) = \bigl((\theta + \Delta\theta) \bmod \pi\bigr) - \dfrac{\pi}{2}$$

(11)

$$\Delta\theta = \operatorname{atan}\left( \dfrac{u - \left\lceil \dfrac{X-1}{2} \right\rceil}{v - \left\lceil \dfrac{Y-1}{2} \right\rceil} \right),$$

(12)

where $\lceil \sim \rceil$ is the ceiling function, and X, Y are the size of $RF_{SAFT-CF,\theta}(x,y)$ in two dimensions, respectively. Besides, the normalization condition has to be satisfied, as discussed by Jeon *et al.* [14]. Finally, using the inverse Fourier transform, we can transform $K_{merged}$ back to the spatial domain, which is the final result after applying D-SAFT and is denoted as $RF_{D-SAFT}(x,y)$.

*2.3. R-L deconvolution and MB deconvolution*

As mentioned previously, acoustic diffraction still limits lateral resolution to no better than in-focus lateral resolution. To further enhance resolution, one can use deconvolution algorithm. Overall, R-L deconvolution and MB deconvolution are two typical methods. The underlying assumption is that an AR-PAM imaging system is a linear and spatial shift-invariant system. Therefore, the acquired PA image, $RF(x,y)$, by AR-PAM can be expressed as:

$$RF(x,y) = p(x,y) \otimes O(x,y) + n(x,y),$$

(13)

where $p(x,y)$ is the PSF of the imaging system, $O(x,y)$ is the original PA source distribution, and $n(x,y)$ is the potential background noise. $\otimes$ is the convolution operator. $p(x,y)$ is the prior knowledge that will be used in the algorithm to achieve resolution enhancement, and $p(x,y)$ is usually obtained from either simulation or experimental measurement. When the PSF is known, R-L deconvolution can be applied iteratively to estimate the original PA source distribution, which is expressed as [23,24]:

$$O'_{i+1}(x,y) = \left[ \dfrac{RF(x,y)}{p(x,y) \otimes O'_i(x,y)} \otimes p(-x,-y) \right] O'_i(x,y),$$

(14)

where $O'_i(x,y)$ is the estimation of $O(x,y)$ in the $i^{th}$ iteration. Different from R-L deconvolution, MB deconvolution expresses the imaging procedure using the following equation:

$$RF(x,y) = H(x,y) * O(x,y) + n(x,y).$$

(15)

where $H(x,y)$ is the dictionary matrix [20,21], and $*$ is the matrix multiplication operator. Therefore, the original PA source distribution could be estimated by the following optimization process:

$$\widehat{O}(x,y) = \underset{O(x,y)}{\operatorname{argmin}} \|RF(x,y) - H(x,y) * O(x,y)\|_2^2,$$

(16)

where $\hat{O}(x,y)$ is the estimation of $O(x,y)$ by minimizing the above residual term. Besides, a regularization term can be used to guarantee sparsity for resolution enhancement [20,21,25,26], and Eq. (16) can be modified as:

$$\hat{O}(x,y) = \underset{O(x,y)}{\mathrm{argmin}} \left( \frac{1}{2} \|RF(x,y) - H(x,y) * O(x,y)\|_2^2 + \lambda \|O(x,y)\|_1 \right),$$

(17)

where $\lambda$ is a parameter to control the degree of sparsity and is usually chosen heuristically. This minimization process can be solved with a fast iterative shrinkage-thresholding algorithm (FISTA implementation in Matlab, https://github.com/tiepvupsu/FISTA) [27]. As optimization of $\hat{O}(x,y)$ in Eq. (17) aims at giving point-like results, Gaussian smoothing always follows [20,21].

## 3. Methods

Overall, our algorithm has two parts, FA-SAFT and D-MB deconvolution, as shown in the green and blue parts in Fig. 2, respectively. They are introduced in Sections 3.1 and 3.2 in the following.

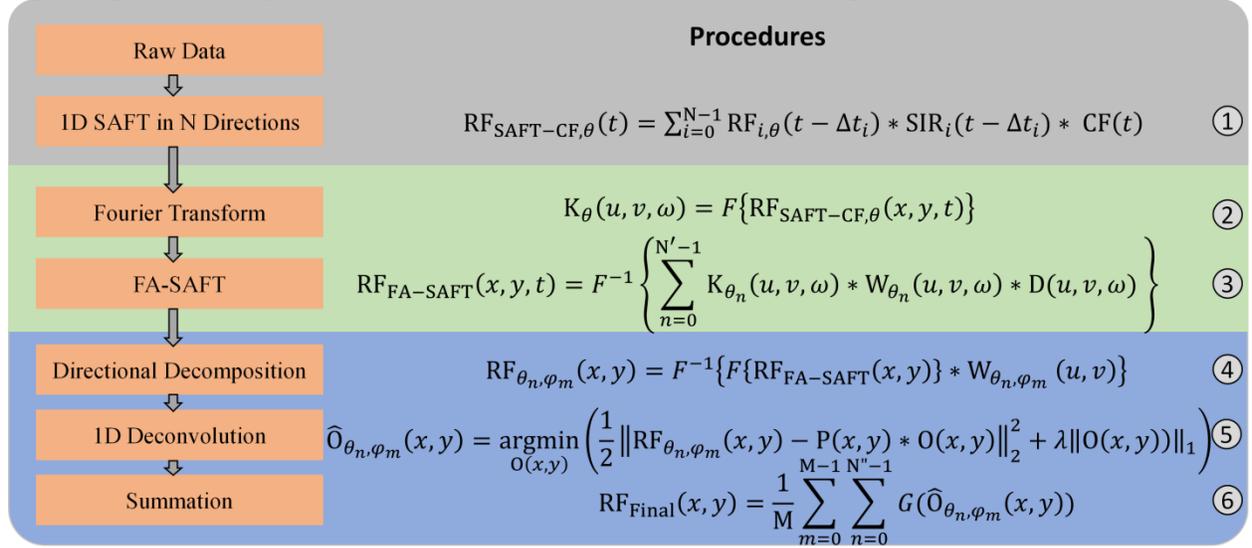

Fig. 2. Procedures of the proposed algorithm.

### 3.1. FA-SAFT

Inspired from the work done by Delbracio et al. [28], we propose to improve resolution and SNR with FA-SAFT. FA-SAFT is modified from D-SAFT. Compared with D-SAFT, FA-SAFT exploits the Fourier bursts among 1D SAFTs along different directions to adaptively enhance the spectrum (k-space), and thus, FA-SAFT can produce improved results.

Note that using CF in SAFT (same for D-SAFT and FA-SAFT) can already increase SNR along with its spatial resolution. The key to optimizing image enhancement is to identify the optimal synthesis direction among 1D SAFTs along different directions. Although 1D SAFTs along different directions are utilized in both D-SAFT and FA-SAFT, FA-SAFT adopts an improved approach to combine the 1D SAFTs from all directions, leading to further improved resolution and SNR. Detailed comparison is elaborated in Section 5. FA-SAFT includes the following steps. First, 1D SAFT is applied to a series of directions using Eqs. (6) and (7). Secondly, the signal is transformed into k-space using Eq. (8). Thirdly, $K_\theta(u,v,\omega)$ is accumulated according to the following equations:

$$K_{merged-FA}(u,v,\omega) = \sum_{n=0}^{N'-1} K_{\theta_n}(u,v,\omega) * W_{\theta_n}(u,v,\omega) * D(u,v,\omega)$$

(18)

$$D(u,v,\omega) = \frac{1}{\sum_{n=0}^{N'-1} \mathrm{abs}(K_{\theta_n}(u,v,\omega))^\gamma},$$

(19)

where $D(u,v,\omega)$ is a factor used for signal sharpening, abs is the absolute function, and $\gamma$ is a variable to control the degree of sharpening. When $\gamma$ is set to 0, FA-SAFT will be reduced to D-SAFT. Since the spectrum of $K_{\theta_n}(u,v,\omega)$ has higher values in the low-frequency range, its reciprocal, $D(u,v,\omega)$, has lower values in the low-frequency range but higher values in the high-frequency range. That is, $D(u, v, \omega)$ can increase the high-frequency components in k-space and thus can enhance resolution (details described in Section 5). Similarly, using the inverse Fourier transform, we can transform $K_{\text{merged-FA}}$ back to spatial domain to get the results after applying FA-SAFT and is denoted as $RF_{\text{FA-SAFT}}(x, y, t)$.

Note that the proposed FA-SAFT approach can be further divided into 2 types, depending on whether the directional term $W_{\theta_n}(u, v, \omega)$ is applied in Eq. (18) or not. The 2 types are termed FA-SAFT-dir0 and FA-SAFT-dir1, which are the cases of without applying the directional term and with applying it in Eq. (18), respectively. Then, the results after FA-SAFT will be used for deconvolution.

*3.2. D-MB deconvolution*

As the line pattern only has 1D sparsity, it stands to reason to apply the 1D deconvolution. Therefore, it is crucial to recognize the branches in different directions so that the 1D deconvolution along different directions can be applied accordingly. We use the hamming window methods as before (Eq. (10)) to decompose the branches into two directions that are perpendicular to each other and apply the 1D MB deconvolution with sparsity (Eq. (17)) for the branches. The algorithm is thus termed D-MB deconvolution. The detailed procedure is described in the following equation:

$$RF_{\theta_n, \varphi_m}(x, y) = F^{-1}\{F\{RF_{\text{FA-SAFT}}(x, y)\} * W_{\theta_n, \varphi_m}(u, v)\}, \tag{20}$$

where $W_{\theta_n, \varphi_m}(u, v)$ has the same definition as in Eq. (10), except the phase shift $\varphi_m$, which is associated with image rotation. $\varphi_m$ is added to Eq. (20) to better extract the signals from the line pattern along different directions over the lateral plane. The modified $W_{\theta_n, \varphi_m}(u, v)$ from Eq. (10) is defined in the following:

$$\overline{\theta_n}(u, v) = \left((\theta + \Delta\theta) \bmod \pi\right) - \frac{\pi}{2} + \varphi_m \tag{21}$$

$$\varphi_m = \frac{m\pi}{2M}, 0 \leq m \leq M - 1, \tag{22}$$

where $\Delta\theta$ is the same as Eq. (12), and M is the number of times to apply deconvolution for different $\varphi_m$. Accordingly, the 1D MB deconvolution with sparsity applied to $RF_{\theta_n, \varphi_m}(x, y)$ along a certain direction $\theta_n$ and a certain $\varphi_m$ can be easily modified from (17) and is expressed as:

$$\hat{O}_{\theta_n, \varphi_m}(x, y) = \underset{O(x, y)}{\text{argmin}} \left(\frac{1}{2} \|RF_{\theta_n, \varphi_m}(x, y) - P(x, y) * O(x, y)\|_2^2 + \lambda \|O(x, y)\|_1\right). \tag{23}$$

Note that $\theta_n$ indicates the deconvolution direction, which lies in the lateral plane and means the direction along which the 1D deconvolution is conducted. The final processed result after D-MB deconvolution is obtained by:

$$RF_{\text{Final}}(x, y) = \frac{1}{M} \sum_{m=0}^{M-1} \sum_{n=0}^{N''-1} G(\hat{O}_{\theta_n, \varphi_m}(x, y)), \tag{24}$$

where N" is set to two. As mentioned above, the hamming window methods are used to decompose the branches into two perpendicular directions (i.e., N" = 2), and 1D MB deconvolution is then applied for the branches. $G$ represents the Gaussian smoothing operator following previous practice [20,21] and was applied during composition. Note that since D-MB deconvolution is applied to 2D images, it can be applied to a lateral maximum amplitude projection (MAP) image or 2D lateral slices at each axial position.

FA-SAFT can be considered to be independent of D-MB deconvolution. That is, FA-SAFT and D-MB deconvolution (with well-estimated PSF) can be applied alone. Therefore, N′ in (18) is independent of N" in (24), and N′ can be different from N". Note that N" = 2 is used because we found that too large N" leads to too straight branches and thus loses fidelity.

## 4. Results

We demonstrated the feasibility and advantages of our algorithm through imaging experiments of tungsten wire phantom, leaf skeleton phantom, and mouse dorsal and ear blood vessels *in vivo*. These experiments were conducted using a focused US transducer with center frequency of 50 MHz, NA of 0.44, and focal length of 6.7 mm. The results are described in the following subsections. The number of iterations for R-L deconvolution (used for comparison) is 15 for tungsten wire experiments and 10 for leaf skeleton experiment and *in vivo* mouse experiments.

*4.1. Tungsten wire phantom imaging*

We designed two tungsten wire phantom imaging experiments to elaborate the effect of our algorithm. First, we imaged two crossed tungsten wires with an angle of ~90° to quantify full width at half maximum (FWHM) and DOF. Secondly, we imaged two closely-located tungsten wires to calibrate the minimum resolvable distance.

*4.1.1. Crossed tungsten wires*
Two tungsten wires with diameter of 20 μm were evenly placed at four different OFDs of 0 mm, 0.3 mm, 0.6 mm and 0.9 mm (i.e., the axial distance of 6.7 mm, 7.0 mm, 7.3 mm and 7.6 mm, respectively, from the surface of the transducer), with one wire overlapping on the other. The four OFDs are denoted as D1–D4, respectively. The lateral MAP images of raw data (i.e., $RF(x,y)$ without further processing) are shown in Supplementary Fig. S1. Supplementary materials are available online. In the images in Supplementary Fig. S1, lateral resolution deteriorates as the abs(OFD) of the wires increases. For convenient comparison of several algorithms, we selected the same regions with the same size of the images for display (indicated by the white boxes in Supplementary Fig. S1).

The workflow of our algorithm is shown in Fig. 3. PAM images of D2 are used for illustration. The $\gamma$ used in FA-SAFT is set to 0.2 (Eq. (19)). We also compared FA-SAFT using different $\gamma$ (see Supplementary Fig. S2). Larger $\gamma$ generates smaller features size (or FWHM), but too large $\gamma$ leads to artifacts. As a result, the optimal $\gamma$ is set to 0.2. For Fig. 3, $N' = 2$ and $M = 1$ are used for illustration. Note that $N' = 16$ (as recommended previously [14]) and $M = 4$ are used for other figures unless otherwise specified. $M = 4$ is chosen to balance the processed effects and the computation time (see Supplementary Fig. S3). Supplementary Fig. S3 shows that the initial phase shift $\varphi_0$ does not affect the D-MB results for the cases of $M = 4$ and $M = 8$. Experimentally measured PSF of 65 μm is used when applying the deconvolution algorithm [13]. Note that since FA-SAFT is to restore the lateral resolution in the out-of-focus region to the in-focus lateral resolution, the same lateral PSF can be assumed for AR-PAM images at different OFDs when applying deconvolution [13].

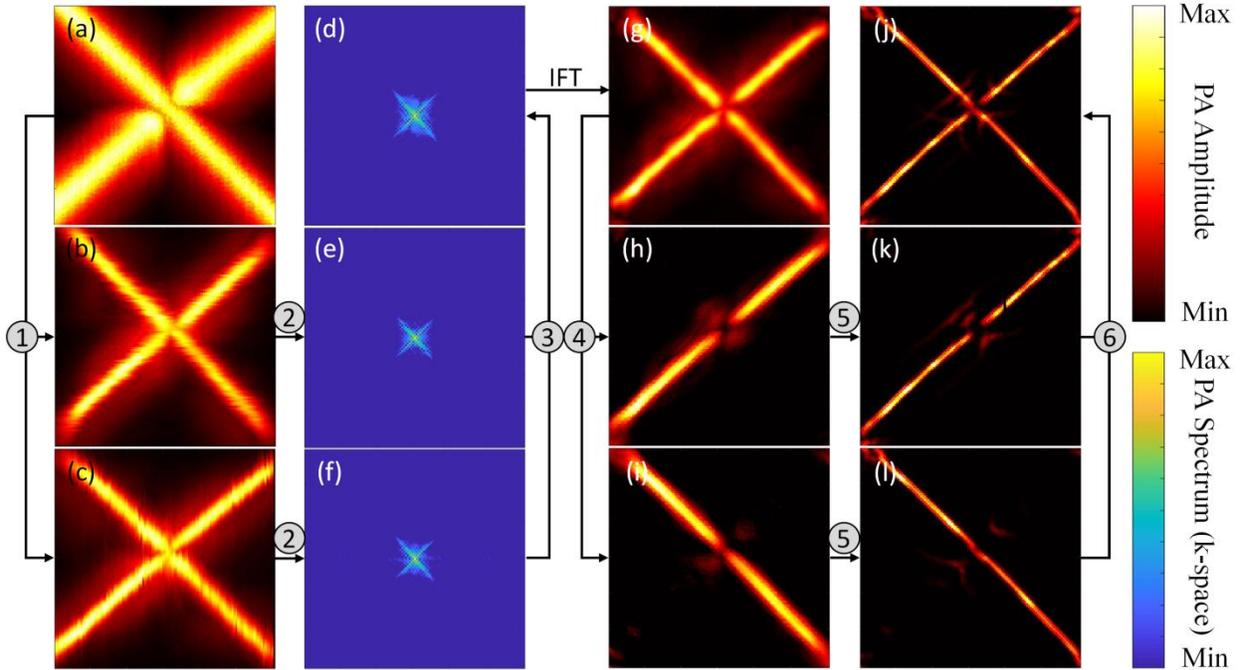

Fig. 3. Illustration of the workflow of our algorithm. The numbers inside circles here correspond to the 6 steps in Fig. 2. $N' = 2$ and $M = 1$ are used here for illustration. (a) MAP image of raw data at D2 (OFD of 0.3 mm). (b,c) By 1D SAFT along the horizontal and vertical directions, respectively. (d–f) (e) and (f) are the Fourier transform of (b) and (c), respectively. (d) is the merged result from (e) and (f). That is, (d) is $K_{merged-FA}(u,v)$. (g) By FA-SAFT (i.e., $RF_{FA-SAFT}(x,y)$). (h,i) The images in spatial domain after directional decomposition of (g). (j–l) (k) and (l) are 1D MB deconvolution results of (h) and (i), respectively. (j) is the final result (i.e., $RF_{Final}(x,y)$). IFT, inverse Fourier transform.

We then studied the effects of the directional term (Eq. (18) with and without $W_{\theta_n}(u,v,\omega)$) and sharpening term (Eq. (18) with and without $D(u,v,\omega)$) in FA-SAFT. PAM images of D2 are used for illustration. The results are shown in Fig. 4. Figs. 4a–4d show the results as follows: without $W_{\theta_n}(u,v,\omega)$ and without $D(u,v,\omega)$ for Fig.

4a (termed pure SAFT); with $W_{\theta_n}(u,v,\omega)$ and without $D(u,v,\omega)$ for Fig. 4b (i.e., D-SAFT); without $W_{\theta_n}(u,v,\omega)$ and with $D(u,v,\omega)$ for Fig. 4c (i.e., FA-SAFT-dir0); with $W_{\theta_n}(u,v,\omega)$ and with $D(u,v,\omega)$ for Fig. 4d (i.e., FA-SAFT-dir1). We found that Fig. 4d achieves better resolution than Fig. 4b. We also compared the performance of R-L deconvolution and D-MB deconvolution based on Figs. 4b and 4d. Figs. 4e and 4f are the results by R-L deconvolution from Figs. 4b and 4d, respectively, and Figs. 4g and 4h are the results by D-MB deconvolution from Fig. 4b and 4d, respectively. The results show that Fig. 4f (finer feature size and fewer artifacts) is better than Fig. 4e, showing the advantage of the proposed FA-SAFT over D-SAFT for the following deconvolution. Similarly, Fig. 4h shows slightly smaller feature size than Fig. 4g. Further, Fig. 4h shows smaller feature size, higher SNR, and fewer artifacts than Fig. 4f, indicating the benefit of D-MB deconvolution over R-L deconvolution. Moreover, we compared the deconvolution results based on Figs. 4c and 4d (see Supplementary Fig. S4). The results show that Supplementary Fig. S4f has comparable resolution to Supplementary Fig. S4c (similarly, Supplementary Fig. S4e comparable to Supplementary Fig. S4b), meaning that both FA-SAFT-dir0 and FA-SAFT-dir1 suit for the following deconvolution. Overall, for SAFT, FA-SAFT outperforms D-SAFT. As for deconvolution, D-MB deconvolution presents better results than R-L deconvolution.

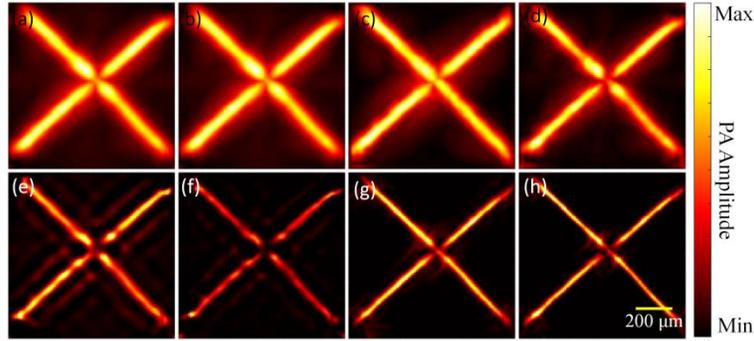

Fig. 4. Comparison of different SAFT and the following deconvolution methods. (a) By pure SAFT. (b) By D-SAFT. (c) By FA-SAFT-dir0. (d) By FA-SAFT-dir1. (e,f) R-L deconvolution from (b) and (d), respectively. (g,h) D-MB deconvolution from (b) and (d), respectively.

Then, we compared the results by FA-SAFT-dir1 and D-SAFT in the out-of-focus region (D2−D4). Besides, R-L deconvolution and D-MB deconvolution were applied and compared for the four OFDs (D1−D4). The results are shown in Fig. 5, which are displayed in MAP images. Fig. 5a shows the MAP images of raw data at D1−D4. Figs. 5b and 5c are the results by D-SAFT and FA-SAFT-dir1, respectively. Figs. 5d−5f show the results by R-L deconvolution, 2D MB deconvolution (Eq. (17)), and D-MB deconvolution from Fig. 5c, respectively. The FWHM and SNR of the images in Fig. 5 are summarized in Table 1. Note that "R-L", "MB", and "D-MB" in Table 1 represent the deconvolution results from Fig. 5c (FA-SAFT-dir1). Overall, the results in Fig. 5 and Table 1 for the results at D1−D4 are consistent with Fig. 4. First, FA-SAFT-dir1 performs better than D-SAFT. The former has smaller FWHM and higher SNR. Secondly, compared with R-L deconvolution, D-MB deconvolution achieves better results with smaller FWHM, higher SNR, better pattern continuity, and fewer artifacts. Compared with MB deconvolution, D-MB deconvolution keeps the previous advantages except for larger FWHM at D2 and D3. Finally, we obtained FWHMs of 26−31 μm within DOF of 1.8 mm [= 0.9 mm×2], when considering below and above the focal plane, by D-MB deconvolution from FA-SAFT images.

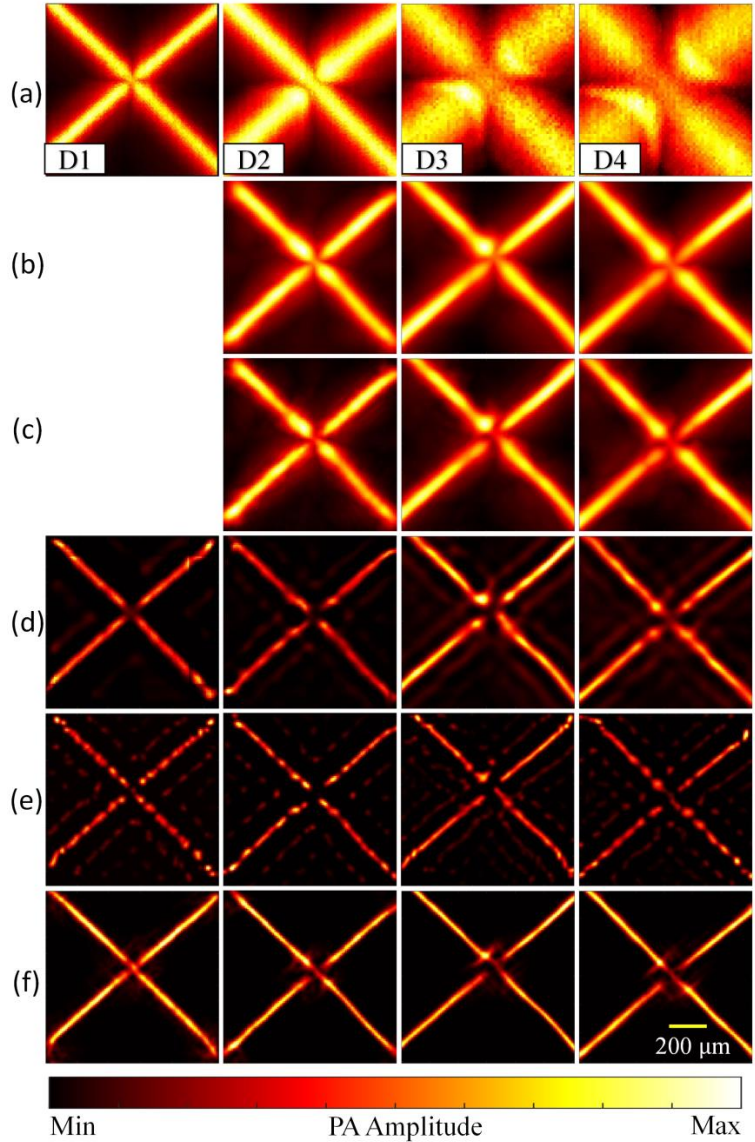

Fig. 5. Results of the first tungsten wire experiment. (a) MAP images of raw data at D1–D4. (b) By D-SAFT. (c) By FA-SAFT-dir1. (d) R-L deconvolution from (c). (e) 2D MB deconvolution from (c). (f) D-MB deconvolution from (c).

Table 1. FHWM (μm) and SNR (dB)

| OFD (mm) | 0 | | 0.3 | | 0.6 | | 0.9 | |
|---|---|---|---|---|---|---|---|---|
| | FWHM | SNR | FWHM | SNR | FWHM | SNR | FWHM | SNR |
| Raw | 64.0 | 21.0 | 152.0 | 23.4 | 311.0 | 17.2 | 426.0 | 14.9 |
| D-SAFT | / | / | 69.0 | 31.6 | 85.0 | 28.4 | 92.0 | 39.5 |
| FA-SAFT | / | / | 61.0 | 33.9 | 68.0 | 33.4 | 77.0 | 43.9 |
| R-L | 36.0 | 35.3 | 39.0 | 35.9 | 33.0 | 37.4 | 39.0 | 46.6 |
| MB | 32.0 | 29.0 | 24.0 | 34.9 | 25.0 | 37.3 | 37.0 | 42.7 |
| D-MB | 28.0 | 47.1 | 26.0 | 42.2 | 27.0 | 51.1 | 31.0 | 57.9 |

## 4.1.2. Closely-located tungsten wires

Two wires with diameter of 20 μm were closely placed with a narrow angle between them. The phantom was imaged at OFDs of 0 mm (i.e., in-focus) and −0.45 mm (i.e., out-of-focus), respectively. The results are shown in Fig. 6. PAM images of raw data at the focal and out-of-focus regions are shown in Figs. 6a and 6d, respectively. For the in-focus image, we directly applied R-L deconvolution (Fig. 6b) and D-MB deconvolution (Fig. 6c), respectively. For the out-of-focus image, we first applied FA-SAFT-dir1 (Fig. 6e). Then, Fig. 6e was further processed by R-L deconvolution (Fig. 6f) and D-MB deconvolution (Fig. 6g), respectively.

The resolution is determined by checking the minimum distinguishable separation (i.e., minimum resolvable distance) of the two tungsten wires (≥6 dB for the contrast between the minimum of the two peaks and the valley). Figs. 6h−6k show 1D profiles along the lines #1−#4 in Fig. 6, respectively (#1 and #2 at the same positions for Figs. 6a−6c; #3 and #4 at the same positions for Figs. 6d−6g). Note that lines #1−#4 in Fig. 6 are chosen based on the above criterion to determine the resolution of Figs. 6c, 6a, 6g, and 6e, respectively. Similarly, the resolution of Figs. 6b, 6d, and 6f can be determined using the same criterion (related 1D profiles not shown). As a result, the measured resolutions of Figs. 6a−6g are 129 μm, 76 μm, 49 μm, 406 μm, 108 μm, 65 μm, and 46 μm, respectively. For OFD of 0 mm, the resolution is enhanced from 129 μm to 49 μm (~2.6 times) before and after applying D-MB deconvolution (Fig. 6a vs. Fig. 6c). For OFD of −0.45 mm, the resolution is enhanced from 108 μm to 46 μm (~2.3 times) before and after applying D-MB deconvolution (Fig. 6e vs. Fig. 6g). Besides, the resolution achieved by D-MB deconvolution is also better than that by R-L deconvolution (~1.4−1.6 times improvement).

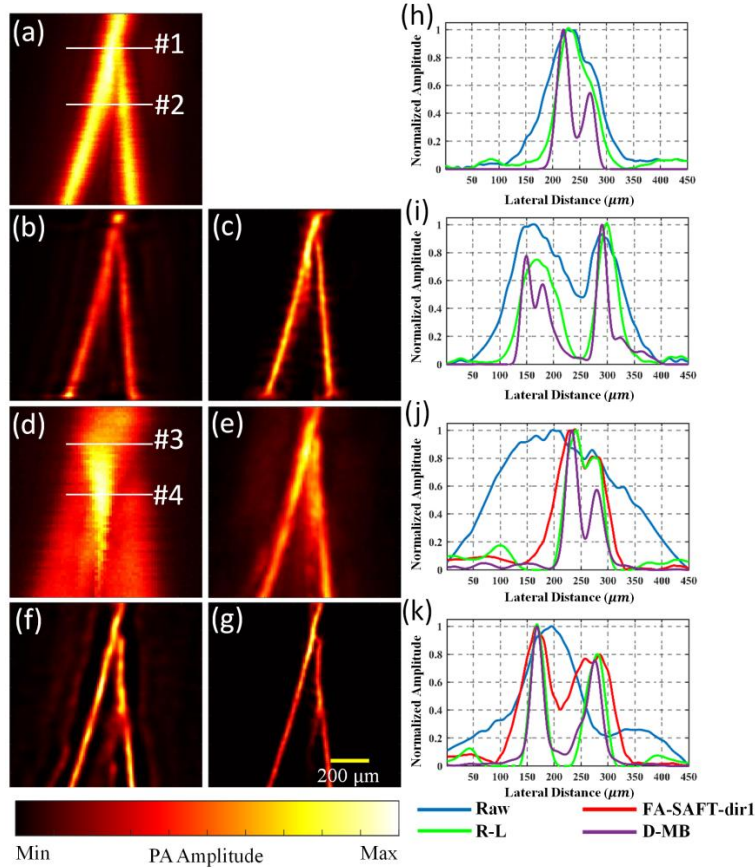

Fig. 6. Results of the second tungsten wire experiment. (a) PAM image of raw data at OFD of 0 mm. (b,c) R-L deconvolution and D-MB deconvolution from (a), respectively. (d) PAM image of raw data at OFD of −0.45 mm. (e) By FA-SAFT-dir1 from (d). (f,g) R-L deconvolution and D-MB deconvolution from (e), respectively. (h,i) 1D profiles of (a−c) along the two lines indicated by #1 and #2 in (a), respectively. (j,k) 1D profiles of (d)-(g) along the two lines indicated by #3 and #4 in (d), respectively. #1-#4: The position below which the two tungsten wires start to be distinguishable in (c), (a), (g), and (e), respectively. That is, (h−k) show the resolution of (c), (a), (g), and (e), respectively.

*4.2. Leaf skeleton phantom imaging*

To investigate the ability of our algorithm for more complex structures, we imaged a Banyan leaf skeleton phantom and applied our algorithm. Before imaging, the leaf was immersed in carbon ink to increase light absorption of the phantom. The results are shown in Fig. 7. PAM images were acquired at both the focal (Fig. 7a) and out-of-focus (Fig. 7d) regions. For the in-focus image, we applied the deconvolution algorithms directly (Figs. 7b and 7c). For the out-of-focus image, different SAFT and the following deconvolution were applied (Figs. 7e–7i).

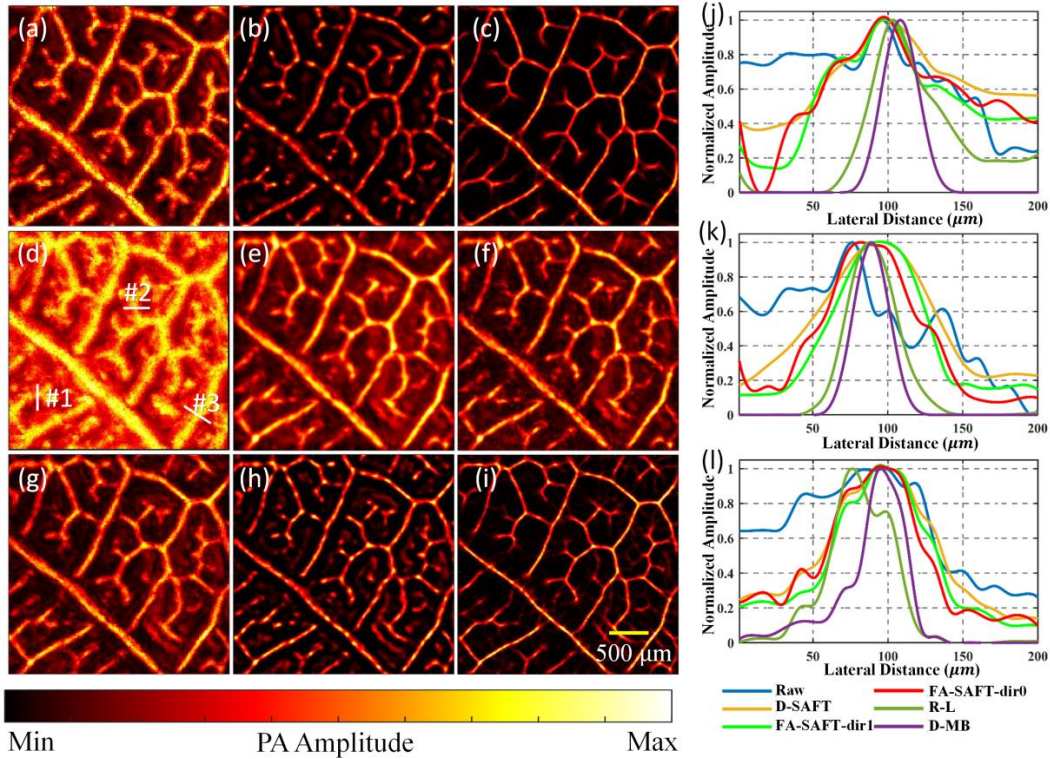

Fig. 7. Results of leaf skeleton experiment. (a) PAM image of raw data at OFD of 0 mm (in-focus). (b,c) R-L deconvolution and D-MB deconvolution from (a), respectively. (d) PAM image of raw data at OFD of 0.25 mm (out-of-focus). (e–g) By D-SAFT, FA-SAFT-dir1, and FA-SAFT-dir0, respectively. (h,i) R-L deconvolution and D-MB deconvolution from (g), respectively. (j–l) 1D profiles of the images (d–i) along the three lines indicated by #1, #2, and #3 in (d), respectively.

First, Fig. 7c presents much better image fidelity and smaller feature size than Fig. 7b, demonstrating the great performance of D-MB deconvolution over R-L deconvolution. Secondly, Figs. 7e–7g are the results by D-SAFT, FA-SAFT-dir1, and FA-SAFT-dir0, respectively. The latter two images show less noise than the former. Similar to Supplementary Fig. S4, FA-SAFT-dir0 (Fig. 7g) is comparable to FA-SAFT-dir1 (Fig. 7f). Therefore, we chose to apply the deconvolution algorithms to the image by FA-SAFT-dir0 (Fig. 7g). The results are shown in Figs. 7h (R-L deconvolution) and 7i (D-MB deconvolution). Similar to Figs. 7b and 7c, Fig. 7i shows better image fidelity and smaller feature size than Fig. 7h. We also applied the deconvolution algorithms to the image by FA-SAFT-dir1 (see Supplementary Fig. S5). The comparable deconvolution results from FA-SAFT-dir1 and FA-SAFT-dir0 in Supplementary Fig. S5 are similar to those in Supplementary Fig. S4.

For FWHM comparison, we plot 1D profiles along the three lines with different directions in Figs. 7d–7i. Note that the three lines are at the same positions in Figs. 7d–7i. For convenience, the three lines are only indicated in Fig. 7d and denoted as #1, #2, and #3. The 1D profiles are shown in Fig. 7j–7l for lines #1, #2, and #3, respectively. FA-SAFT-dir0 and FA-SAFT-dir1 achieve better FWHM than D-SAFT. Besides, as expected, D-MB deconvolution produced smaller FWHM compared with R-L deconvolution. In addition, we rotated Fig. 7(a) by 45° to test the robustness of D-MB deconvolution to different orientations of the line pattern (see Supplementary Fig. S6).

*4.3. In vivo mouse imaging*

We further demonstrated our algorithm applied to *in vivo* mouse imaging. This animal experiment was conducted in conformity with the laboratory animal protocol approved by Laboratory Animal Care Committee of Shanghai Jiao Tong University. Mouse dorsal subcutaneous vessels of a BALB/c mouse (~20 g, Slac Laboratory Animals) with hair removed were first imaged by our AR-PAM system.

Similar to Figs. 7d−7i, the images of raw data (at OFD of ~0.4 mm), by D-SAFT, by FA-SAFT-dir1, by FA-SAFT-dir0, R-L deconvolution after FA-SFAT-dir0, and D-MB deconvolution after FA-SAFT-dir0 are shown in Figs. 8a−8f, respectively. Besides, for resolution comparison, we also plot 1D profiles in Figs. 8g−8i along the three lines in Figs. 8a−8f. The results are overall similar to Fig. 7. First, FA-SAFT-dir0 (Fig. 8d) and FA-SAFT-dir1 (Fig. 8c) present lower noise than D-SAFT (Fig. 8b). Interestingly, FA-SAFT-dir0 (Fig. 8d) shows slightly higher SNR than FA-SAFT-dir1 (Fig. 8c). Therefore, the deconvolution based on Fig. 8d was conducted, and the results are shown in Figs. 8e and 8f by R-L deconvolution and D-MB deconvolution, respectively. Secondly, as expected, D-MB deconvolution (Fig. 8f) achieves better results than R-L deconvolution in terms of image fidelity and SNR. Similarly, the deconvolution applied to Fig. 8c was also conducted for comparison (see Supplementary Fig. S7). Overall, Supplementary Fig. S7f shows slightly higher SNR than the corresponding Supplementary Fig. S7c (similarly for Supplementary Fig. S7e vs. Fig. S7b), which is consistent with the results before applying deconvolution (Supplementary Fig. S7d vs. Supplementary Fig. S7a). Besides, by checking Figs. 8g−8i, FA-SAFT-dir1 and FA-SAFT-dir0 present higher resolution than D-SAFT, and D-MB deconvolution enables the smallest FWHM. Figs. 8j−8o show zoom-in images of the dashed box regions (indicated in Fig. 8a) in Figs. 8a−8f, respectively. Note that the relatively weak PA amplitude of some small vessels in the dashed box region in Fig. 8f does not mean that these small vessels disappear after applying deconvolution, but is due to the adopted color map for display. As can be seen in Fig. 8o, these small vessels can be clearly visualized with good SNR.

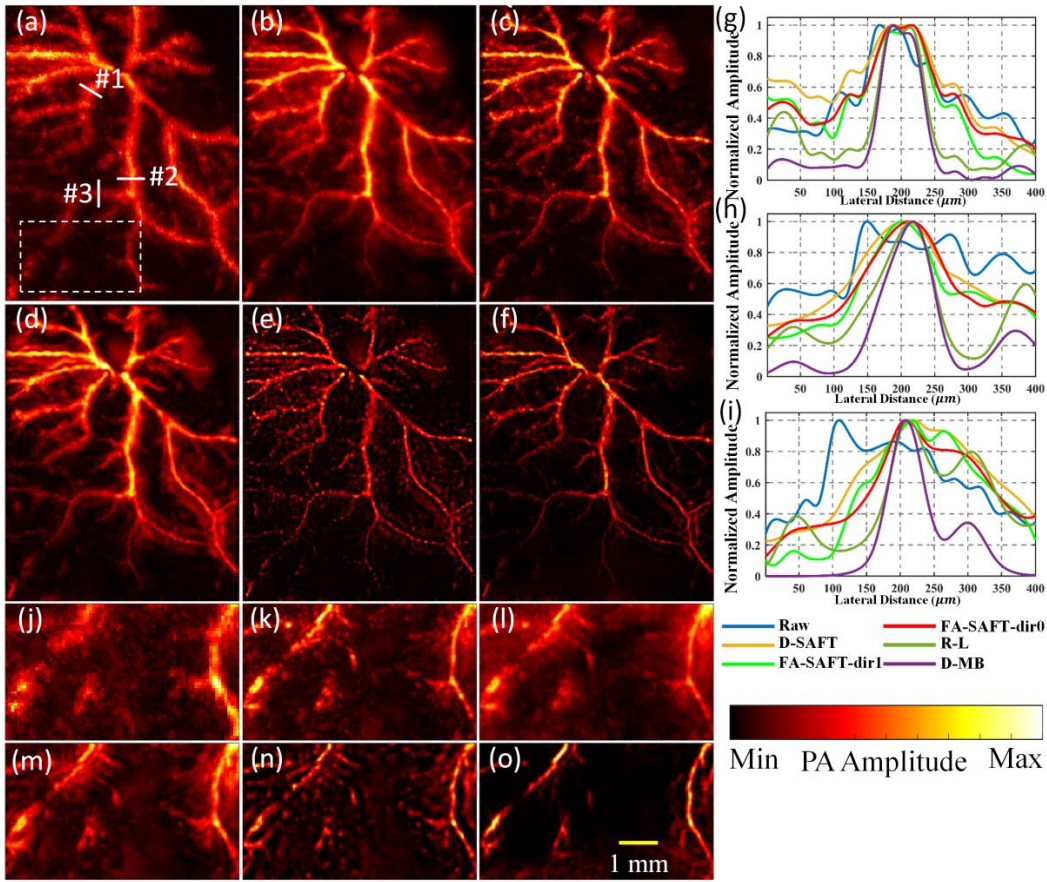

Fig. 8. Results of *in vivo* mouse experiment (mouse dorsal blood vessels). (a) PAM image of raw data at OFD of ~0.4 mm (out-of-focus). (b−d) By D-SAFT, FA-SAFT-dir1, and FA-SAFT-dir0, respectively. (e,f) R-L deconvolution and D-MB deconvolution from (d), respectively. (g−i) 1D profiles of the images (a−f) along the three lines indicated by #1, #2, and #3 in (a), respectively. (j−o) Zoom-in images of the dashed box region (indicated in (a)) in (a−f), respectively. Note that the box indicates the same regions for (a−f).

For further validation, we tested our algorithm on two other *in vivo* AR-PAM data: (i) Another image of mouse dorsal blood vessels *in vivo* [13] (see Supplementary Fig. S8); (ii) mouse ear blood vessels *in vivo* imaged at the focal and out-of-focus regions (Figs. 9 and 10). Fig. 9 shows the images of raw data and processed by SAFT and deconvolution, and Fig. 10 shows 1D profiles and zoom-in images indicated in Fig. 9 for better comparison and display. The results are promising. For Supplementary Fig. S8, overall, FA-SAFT performs better than D-SAFT (notably high resolution in Supplementary Fig. S8d over Fig. S8b), and D-MB deconvolution produces better results than R-L deconvolution (especially pattern continuity in Supplementary Fig. S8f over Fig. S8e). Similarly, for Fig. 9, FA-SAFT enables higher resolution than D-SAFT (FA-SAFT vs. D-SAFT: Figs. 9f vs. 9e). Besides, compared with R-L deconvolution, D-MB deconvolution renders finer feature size (D-MB vs. R-L: Figs. 9c vs. 9b; Figs. 9i vs. 9h) and better pattern continuity (D-MB vs. R-L: Figs. 10f vs. 10e; Figs. 10h vs. 10g). Notably, the high resolution by D-MB deconvolution is clearly evidenced by Figs. 10a−10d, where *D-MB can better resolve two closely-located vessels*. Similar to Fig. 8, the relatively weak PA amplitude of some small vessels in the dashed boxes in Fig. 9 is due to the adopted color map for display. These small vessels can be clearly visualized with good SNR in Figs. 10e−10h.

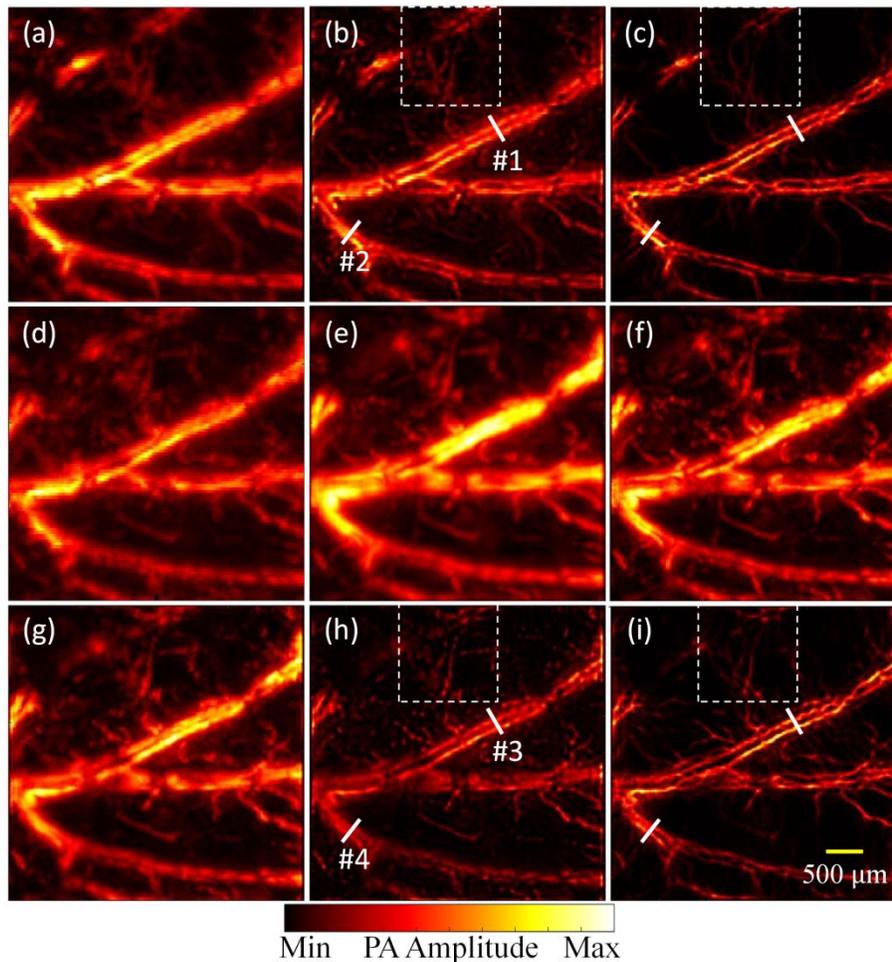

Fig. 9. Results of *in vivo* mouse experiment (mouse ear blood vessels). (a) PAM image of raw data at DOF of 0 mm (in-focus). (b,c) R-L deconvolution and D-MB deconvolution from (a), respectively. (d) PAM image of raw data at DOF of −0.3 mm (out-of-focus). (e−g) By D-SAFT, FA-SAFT-dir1, and FA-SAFT-dir0, respectively. (h,i) R-L deconvolution and D-MB deconvolution from (g), respectively. Note that D-MB deconvolution is suitable for resolving small feature size (e.g., comparable to or smaller than PSF) and has problems in dealing with large feature size. Therefore, (c) and (i) are obtained by scaling down (a) and (g) (i.e., feature size reduced) before applying deconvolution and scaling back (feature size recovered) after deconvolution.

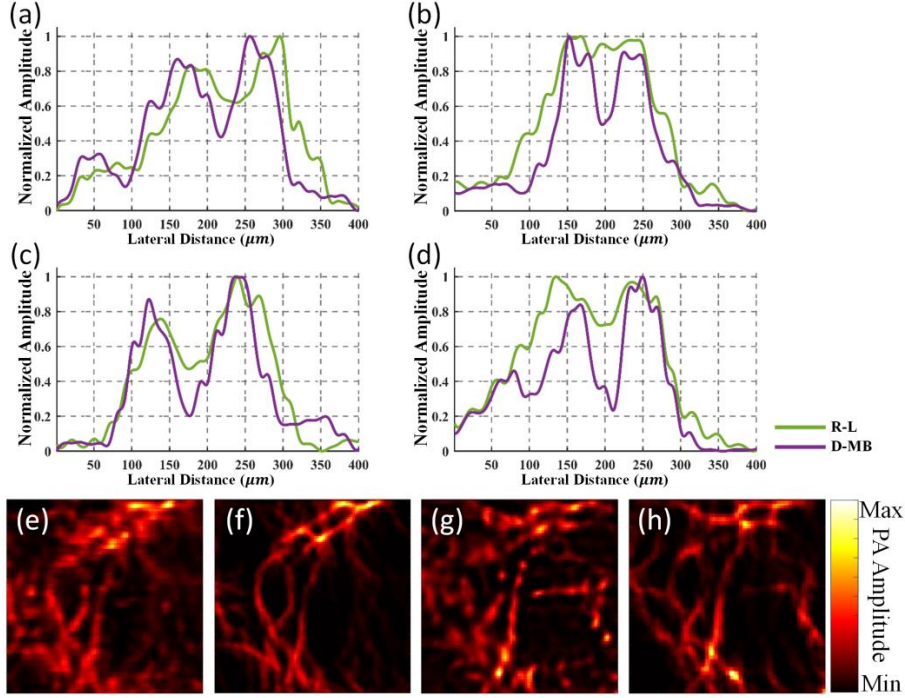

Fig. 10. 1D profiles and zoom-in images for better comparison and display of Fig. 9. (a–d) 1D profiles along the four lines indicated by #1–#4, respectively, in Fig. 9 (#1 and #2 at the same positions for Figs. 9b and 9c; #3 and #4 at the same positions for Figs. 9h and 9i). (e–h) Zoom-in images of Figs. 9b, 9c, 9h, and 9i (dashed boxes), respectively.

## 5. Discussion

We proved that our algorithm can improve the lateral resolution, SNR, and pattern fidelity in both phantom and *in vivo* imaging experiments. The great performance can be attributed to two parts, FA-SAFT and D-MB deconvolution. More details are discussed in the following.

*5.1. FA-SAFT*

FA-SAFT produced higher resolution and sharper edges than D-SAFT. The resolution improvement in FA-SAFT can be explained from the sharpening term, which is newly introduced in FA-SAFT and acts as an adaptive high-pass filter. For better illustration, we compared the processed results and their spectra (k-space) by high-pass filtering (HPF), D-SAFT (i.e., $\gamma = 0$), and FA-SAFT-dir1 of the crossed tungsten wire images over D2–D4 (Section 4.1.1) (see Supplementary Figs. S2 and S9). In Supplementary Fig. S2, we compared the results by HPF using different cut-off frequencies and by FA-SAFT-dir1 using different $\gamma$. We found that HPF with higher cut-off frequency produces smaller feature size, but HPF with too high cut-off frequency of $\geq 3$ units (larger units correspond to higher cut-off frequency) results in evident artifacts. Thus, the optimal cut-off frequency is set to 2 units. As mentioned previously, the optimal $\gamma$ is set to 0.2. Therefore, we then compared the optimized results by HPF with cut-off frequency of 2 units and by FA-SAFT-dir1 with $\gamma = 0.2$. As can be seen, FA-SAFT-dir1 with $\gamma = 0.2$ shows well-suppressed background (or high SNR).

In Supplementary Fig. S9, we first checked the case of D2. Compared with HPF, FA-SAFT-dir1 did not filter out all the low-frequency components but adjusted their weights adaptively, as indicated by the red arrows in Supplementary Fig. S9. Besides, compared with [HPF]–[D-SAFT] ("–" means difference), [FA-SAFT-dir1]–[D-SAFT] can more selectively enhance the images along the directions of the tungsten wires (i.e., increased weights in the spectra of FA-SAFT-dir1 along the two diagonal directions corresponding to the main signals of tungsten wires), showing that the sharpening term in FA-SAFT-dir1 can better reflect directional features of the data compared with HPF. In other words, the sharpening term in FA-SAFT-dir1 can increase the weights of high-frequency components adaptively. Here the "adaptive" means that the sharpening term changes accordingly with the 1D SAFT results as in Eq. (19) and enhances the spectrum directionally. Specifically, for example, as indicated by the green arrows in Supplementary Fig. S9, the spectrum not along the directions of the tungsten wires are more suppressed in [FA-

SAFT-dir1]−[D-SAFT] than in [HPF]−[D-SAFT]. As can be seen in Supplementary Fig. S9, the same characteristics are observed in D3 and D4.

Therefore, Supplementary Fig. S9 would explain that compared with HPF, FA-SAFT-dir1 can enhance resolution (having similar high-frequency components) and without sacrificing SNR (enhancing the spectrum directionally, as detailed in the previous paragraph). Similarly, Supplementary Fig. S9 would also explain compared with D-SAFT, FA-SAFT-dir1 enables higher resolution (keeping more high-frequency components, as shown in the columns of "FA-SAFT-dir1" vs. "D-SAFT") and higher SNR (enhancing the spectrum directionally, as shown in the column of [FA-SAFT-dir1]−[D-SAFT]).

Another variable in FA-SAFT is the number of directions for 1D SAFT ($N'$ in Eq. (18)). As discussed in D-SAFT [14], using more directions (larger $N'$) is useful to improve the resolution of branches in different orientations. However, when $N'$ is large enough ($\geq 16$), the image quality may not improve anymore. Therefore, $N'$ is also set to 16 in our demonstrations.

Overall, FA-SAFT-dir0 produces comparable or slightly better results than FA-SAFT-dir1 (e.g., slightly better for Fig. 8 and Supplementary Fig. S8). Besides, either FA-SAFT-dir0 or FA-SAFT-dir1 performs better than D-SAFT for the demonstrations in this work (Figs. 4–9). This may imply that the sharpening term plays a more important role than the directional term in Eq. (18). On the other hand, the following deconvolution (either R-L or D-MB) exhibits similar trends (e.g., similar results after deconvolution are obtained given similar image quality of FA-SAFT-dir0 and FA-SAFT-dir1).

SAFT in AR-PAM has been continuously improved to deal with more complicated patterns, especially for microvascular patterns, to meet the requirements of *in vivo* imaging applications. By considering 1D sparsity of the line pattern, A-SAFT and D-SAFT were proposed [10,14]. However, the former requires successful recognition of diverse orientations of vessel branches prior to applying A-SAFT, while the latter can better recover the vessel branches by utilizing the weighted 1D SAFTs. Based on D-SAFT, we proposed FA-SAFT to further improve resolution and SNR by introducing the sharpening term, as detailed previously. Deconvolution can perform better given the original input images with better image quality. Taking advantage of the high-quality image by FA-SAFT, D-MB deconvolution using FA-SAFT images outperforms that using D-SAFT images.

*5.2. D-MB deconvolution*

After FA-SAFT, deconvolution algorithms further improve the resolution and SNR. R-L deconvolution algorithm was initially designed for astronomy applications, where the targets are usually point-like objects. Zhu et al. first introduced R-L deconvolution for PAM imaging [29]. Compared to R-L deconvolution and MB deconvolution, the proposed D-MB deconvolution is more suitable for the line pattern because 1D sparsity of the line pattern is considered in D-MB deconvolution. To our knowledge, the directional deconvolution method for handling the line pattern is proposed for the first time. Besides the line pattern, D-MB deconvolution with $M \geq 4$ can well preserve the point features (see Supplementary Fig. S10). Another parameter in D-MB deconvolution is the sparsity regularization term, which contributes to the pattern centralization (i.e., resolution enhancement) and noise reduction to an extent (see Supplementary Fig. S11). Finally, our algorithm (FA-SAFT and D-MB deconvolution) achieved lateral FWHMs of 26−31 μm when imaging 20 μm tungsten wires (Fig. 5), which is better than measured PSF of 65 μm. Besides, lateral resolution of 46−49 μm was obtained when resolving two closely-located tungsten wires (Fig. 6), which is better than original resolution of 129−406 μm. Note that the "sparsity" here does not mean that the pattern of the input PA image has to be sparse in order to apply the D-MB deconvolution. It works well in Figs. 7, 8, and S8, which proves that it is applicable to the AR-PAM images with a certain degree of pattern density.

To improve 3D resolution for the line pattern of blood vessels in AR-PAM images, the proposed D-MB deconvolution can be applied to 2D lateral slices at each axial position, and then 1D R-L or 1D MB deconvolution can be applied along the axial direction [13]. Considering that the line pattern of blood vessels mainly appears over the 2D lateral slices, it would be reasonable to apply 1D deconvolution along the axial direction without dealing with the directional characteristic. Besides, such a method for 3D deconvolution can save computation time and reduce memory usage [13].

*5.3. Other discussion*

As AR-PAM is a 3D imaging modality, we evaluated the axial resolution changes when using our algorithm (see Supplementary Fig. S12 and Supplementary Table S1). We found that axial resolution after FA-SAFT and D-MB deconvolution is comparable or slightly improved. SAFT is typically not expected to improve the axial resolution [13]. On the other hand, in Supplementary Fig. S12, D-MB deconvolution is applied to 2D lateral slices at each axial position (multiple-times deconvolution for many slices). Because deconvolution is not applied along the axial direction, the resolution enhancement along the axial direction is not obvious. Note that except Supplementary Fig. S12, D-MB deconvolution is performed over the MAP images (e.g., Figs. 7i and 8f).

The robustness of the algorithm to different SNR is also an important factor worthy to discuss. Indeed, sufficient SNR is essential for deconvolution. Fortunately, in our algorithm, the first step, FA-SAFT, helps improve SNR for out-of-focus images prior to applying the second step, D-MB deconvolution. In Supplementary Fig. S13 and Supplementary Table S2, the effect of SNR on the resolution enhancement by FA-SAFT and D-MB deconvolution is evaluated. Note that the resolution in Supplementary Table S2 is determined using a criterion similar to Fig. 6. When artifacts become too evident to determine resolution, the resolution is regarded as "not applicable (N/A)." As can be seen, when the PAM image of raw data has peak SNR (PSNR) down to 20 dB (Fig. 6d with noise added), FA-SAFT and D-MB deconvolution can still enhance resolution from N/A to 54 μm. On the other hand, in Supplementary Fig. S14 and Supplementary Table S3, the effect of SNR on the resolution enhancement by D-MB deconvolution alone is investigated. The determination of resolution in Supplementary Table S3 is the same as Supplementary Table S2 mentioned above. As can be seen, when the PAM image by FA-SAFT (prior to applying deconvolution) has PSNR down to 21 dB (Fig. 6e with noise added), D-MB deconvolution can still enhance resolution from N/A to 54 μm, while R-L deconvolution suffers artifacts and the resolution is still N/A. Overall, in Supplementary Fig. S14, D-MB deconvolution produces better image quality (i.e., fewer artifacts) than R-L deconvolution.

Another important factor is the performance of the proposed algorithm when applied to images with penetration depth in tissue. We conducted additional *in vivo* AR-PAM imaging of subcutaneous blood vessels around the lower abdomen region of a mouse. As can be seen in Supplementary Fig. S15, the proposed algorithm demonstrates good performance for the *in vivo* AR-PAM images with penetration depth of ~0.32 mm. Besides, although the skin surface is not flat and the blood vessels distribute over a large axial range, our algorithm applied for a large axial range of 1.35 mm (OFD from −0.525 mm to 0.825 mm) shows satisfactory results. We further and intentionally sought deeper blood vessels in AR-PAM imaging experiments. As can be seen in Supplementary Fig. S16 (notably the blood vessel V3), the proposed algorithm still demonstrates the effectiveness of resolution enhancement for the *in vivo* AR-PAM image of deep vessels up to ~1.3 mm below the skin surface.

In the experiments described above, the penetration depth of ~1.3 mm was demonstrated. AR-PAM is able for deeper penetration in tissue. In this case, however, SNR degrades, and low SNR adversely affects the performance of the proposed algorithm, as described previously (Supplementary Fig. S13, Table S2, Fig. S14, and Table S3). To advance the penetration of *in vivo* AR-PAM when applying our algorithm, one could consider signal averaging and/or contrast agents to secure sufficient SNR, which would ensure *in vivo* applicability of our algorithm.

In Section 4.1, the tungsten wires were immersed in non-scattering water media and placed at different OFDs (e.g., 0–0.9 mm), and the performance of the algorithms at different OFDs can be conveniently compared (Table 1 and Figs. 5 and 6) by excluding the scattering effect. Similar experiments were conducted also in non-scattering water media in a previous study of resolution enhancement in AR-PAM [30]. Similar to the above discussion, strong scattering leads to low SNR, which degrades the performance of the proposed algorithm. Therefore, for scattering media, approaches to boost SNR are needed to ensure the performance of our algorithm.

The computation time of our algorithm is another important aspect. For FA-SAFT, the computation time is mainly determined by the number of directions for 1D SAFT ($N'$) and is proportional to $N'$, which is similar to the case of D-SAFT (Supplementary Table S4). That is, the inclusion of the sharpening term (specifically, for FA-SAFT-dir1) in the calculation will increase the computation time little (the time increase of ~0.2 seconds in Supplementary Table S4 for $N' = 16$). Therefore, once the same $N'$ is chosen for D-SAFT and FA-SAFT, they will have similar computation time (Supplementary Table S4). For D-MB deconvolution, more computation time is required compared with R-L deconvolution (Supplementary Table S5). Fortunately, the computation time of D-MB deconvolution is less than that of FA-SAFT, and thus, the overall computation time of our algorithm is still at the same time scale as FA-SAFT (FA-SAFT of ~72 s ($N' = 16$) and D-MB deconvolution of ~8 s ($M = 4$) in Supplementary Table S4 and Table S5, respectively).

## 6. Conclusions
Tailored for AR-PAM imaging applications of microvasculature, we proposed a novel algorithm consisting of FA-SAFT and D-MB deconvolution for AR-PAM to achieve evident image enhancement in both the focal and out-of-focus regions. Specifically, FA-SAFT improved resolution and SNR, and D-MB deconvolution further enhanced resolution and SNR while keeping image fidelity. As a result, compared with other SAFT and/or deconvolution algorithms, our algorithm achieved great performance in terms of resolution, SNR, and image fidelity. Notably, we demonstrated promising results for in vivo images of mouse blood vessels by our algorithm given the images of raw data with sufficient SNR. Our work opens up new opportunities for microvascular AR-PAM imaging applications.

**Acknowledgments.**
This work was supported by National Natural Science Foundation of China (NSFC) (61775134).

**Declaration of Competing Interest.**
The authors declare no conflicts of interest.

**Data and code availability.**
The data and codes that support the findings of this study are available from the corresponding author upon reasonable request.

**Author Contributions.**
FF, SL, and SLC initiated the project and designed the study. FF developed the algorithm. SL performed experiments. FF and SL analyzed the data. FF and SLC drafted the manuscript. All authors reviewed and revised the manuscript.

**Appendix A.** Supporting information
Supplementary materials associated with this article can be found in the online version at xxxx.

# Supplementary Materials

# Image enhancement in acoustic-resolution photoacoustic microscopy enabled by a novel directional algorithm

Fei Feng+, Siqi Liang+, and Sung-Liang Chen*

**Supplementary Tables**

Table S1. Axial resolution measurement in FWHM (μm)

|              | D1 | D2 | D3 | D4 |
|--------------|----|----|----|----|
| Raw          | 35 | 36 | 37 | 39 |
| D-SAFT       | -  | 34 | 41 | 42 |
| FA-SAFT-dir0 | -  | 30 | 30 | 41 |
| FA-SAFT-dir1 | -  | 30 | 36 | 38 |
| D-MB         | 34 | 31 | 22 | 33 |

The above axial resolutions are obtained from the left objects in Fig. S12 (below).

Table S2. Resolution comparison for Fig. S13

|                  | Reference | PSNR of (a) | | |
|------------------|-----------|-------|-------|-------|
|                  |           | 23 dB | 20 dB | 17 dB |
| (a) Raw data     | 406       | -     | -     | -     |
| (b) FA-SAFT-dir1 | 108       | -     | -     | -     |
| (c) D-MB         | 46        | 56    | 54    | -     |

Unit: μm

Table S3. Resolution comparison for Fig. S14

|                  | Reference | PSNR of (a) | | |
|------------------|-----------|-------|-------|-------|
|                  |           | 21 dB | 17 dB | 14 dB |
| (a) FA-SAFT-dir1 | 108       | -     | -     | -     |
| (b) R-L          | 65        | -     | -     | -     |
| (c) D-MB         | 46        | 54    | -     | -     |

Unit: μm

Table S4. Computation time of different SAFT methods

| Method | 1D SAFTs (N' = 16) | D-SAFT | FA-SAFT-dir0 | FA-SAFT-dir1 |
| --- | --- | --- | --- | --- |
| Time (Second) | 70.96 | 70.96+1.18 | 70.96+**1.07** | 70.96+1.38 |

The above time corresponds to computing 3D PA data with size $121 \times 121 \times 30$. Total computation time for D-SAFT, FA-SAFT-dir0, and FA-SAFT-dir1 is the addition of the time of 1D SAFTs (N' = 16) and the time of Eqs. (9) and (18) (for D-SAFT and FA-SAFT, respectively) with their inverse Fourier transform.

Table S5. Computation time of different deconvolution methods

| Method | 2D MB | R-L | D-MB (M = 1) | D-MB (M = 2) | D-MB (M = 4) | D-MB (M = 8) |
| --- | --- | --- | --- | --- | --- | --- |
| Time (Second) | 59.17 | 0.04 | 1.19 | 3.52 | 8.01 | 17.17 |

The above time corresponds to computing 2D MAP data with size $121 \times 121$. Note that the same computer resource (Intel® Core™ i7-7700 @ 3.6 GHz CPU and 16.0 GB RAM memory) was used for Table S4 and S5 for fair comparison.

**Supplementary Figures**

All supplementary figures are PAM images displayed in MAP except Figs. S9 (spectra) and S12, S15d, S15e, S16a6, S16b6, and S16c6 (B-mode images). These PAM images share the same color bar of PA amplitudes in Fig. S1.

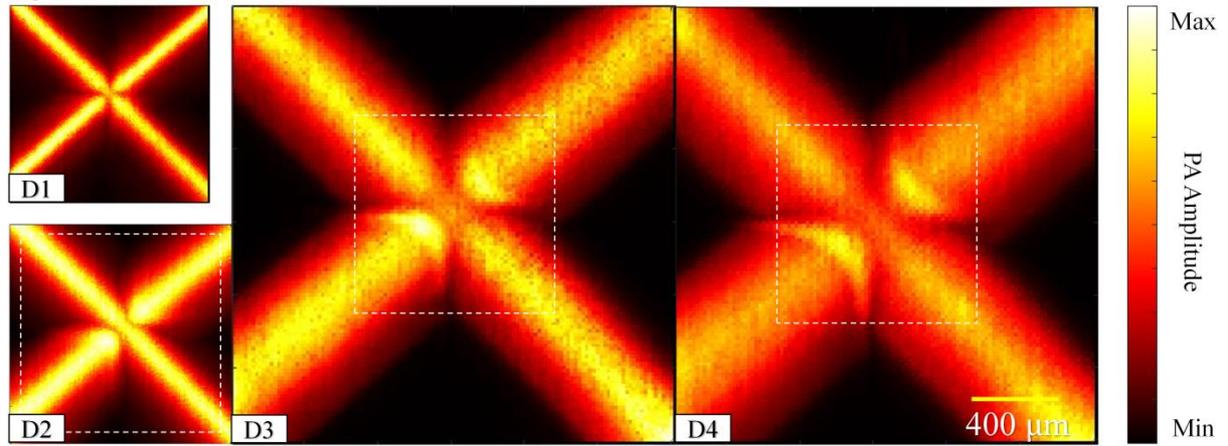

Fig. S1. The lateral MAP images of raw data at OFDs of D1–D4. The white boxes in D2–D4 indicate the same regions with the same size as the image of D1.

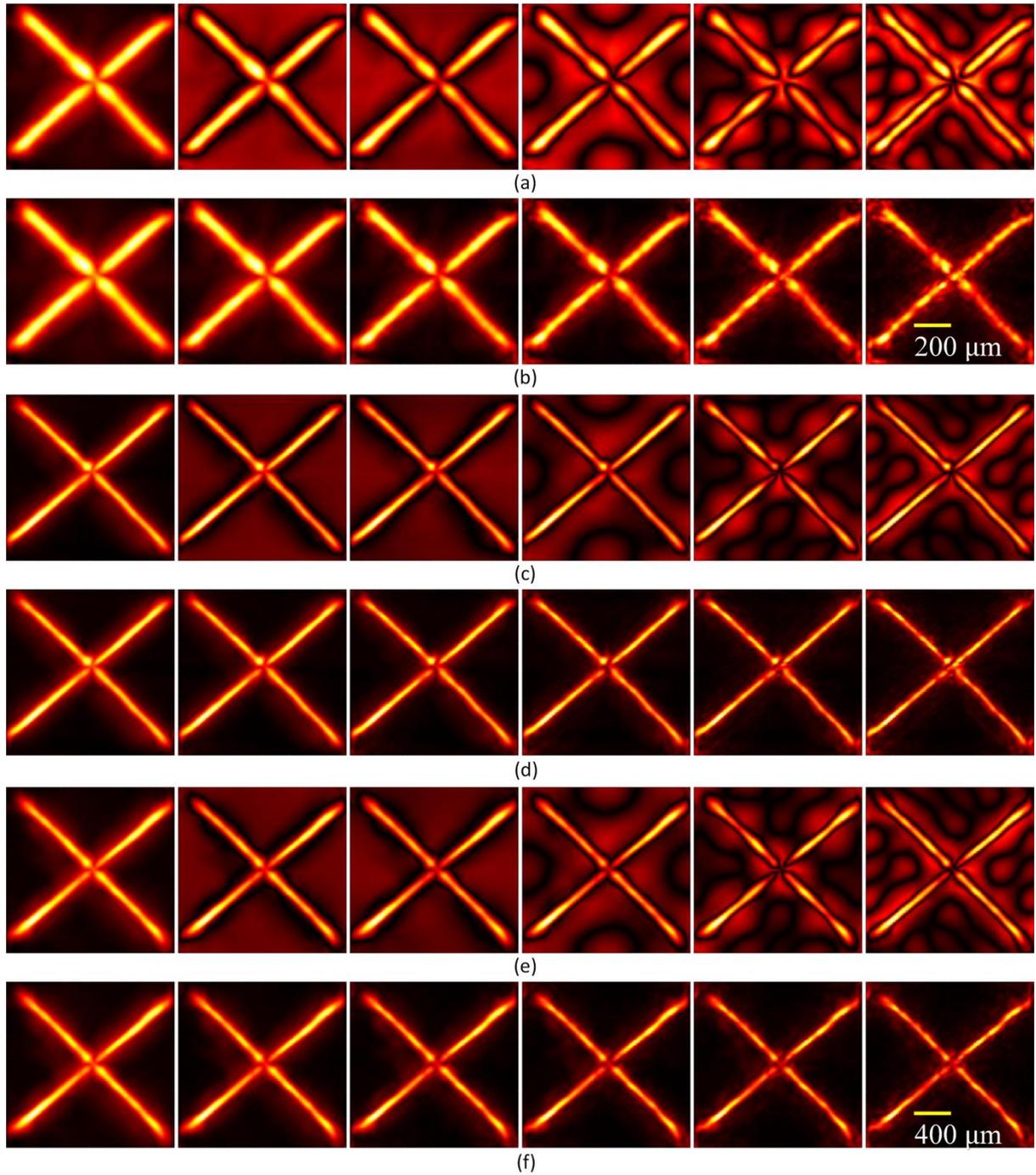

Fig. S2. Comparison of results by HPF and by FA-SAFT-dir1. (a) HPF using cut-off frequencies of 0, 1, 2, 3, 4, and 5 units (from left to right) at D2. (b) FA-SAFT-dir1 using $\gamma = 0$, 0.1, 0.2, 0.3, 0.4, and 0.5 (from left to right) at D2. (c) HPF using cut-off frequencies of 0, 1, 2, 3, 4, and 5 units (from left to right) at D3. (d) FA-SAFT-dir1 using $\gamma = 0$, 0.1, 0.2, 0.3, 0.4, and 0.5 (from left to right) at D3. (e) HPF using cut-off frequencies of 0, 1, 2, 3, 4, and 5 units (from left to right) at D4. (f) FA-SAFT-dir1 using $\gamma = 0$, 0.1, 0.2, 0.3, 0.4, and 0.5 (from left to right) at D4. (a) and (b) share the same scale bar in (b). (c−f) share the same scale bar in (f).

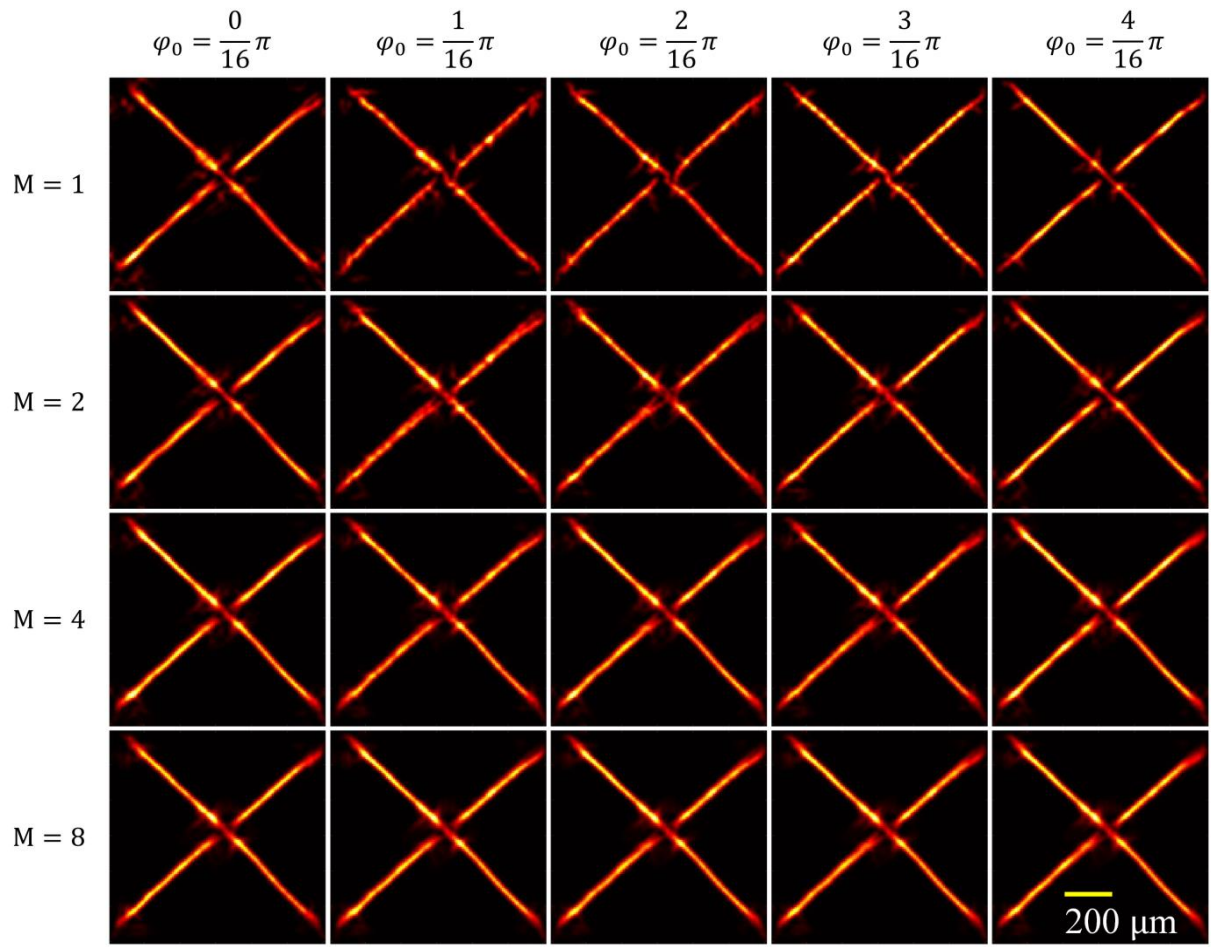

Fig. S3. Results of the crossed tungsten wire phantom at D2 by D-MB deconvolution with various M and $\varphi_0$ in Eq. (24). As can be seen, compared with M = 1, the artificial features can be well suppressed in M = 4 and M = 8. To balance the processed effects and the computation time, M = 4 is chosen for Figs. 4–10.

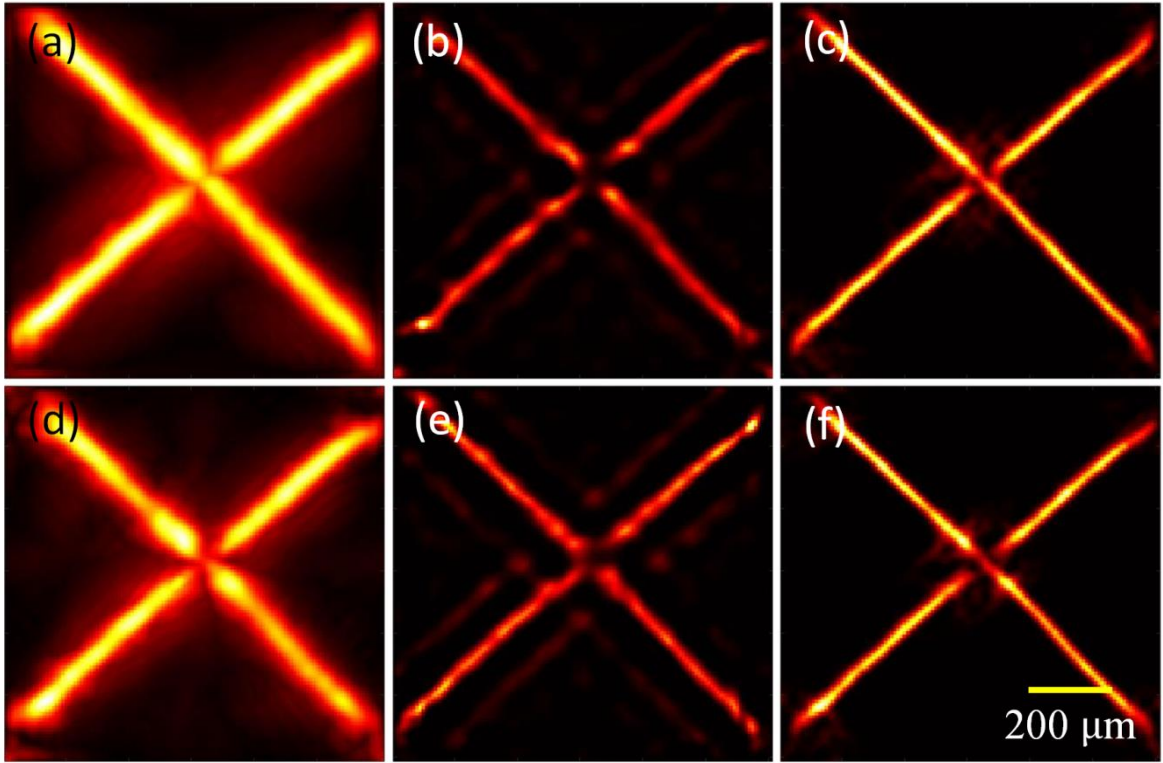

Fig. S4. Comparison of FA-SAFT-dir0 and FA-SAF-dir1 for their following deconvolution results. The crossed tungsten wire phantom at D2 is used. (a) By FA-SAFT-dir0 (same as Fig. 4c). (b,c) R-L deconvolution and D-MB deconvolution from (a), respectively. (d) By FA-SAFT-dir1 (same as Fig. 4d). (e,f) R-L deconvolution and D-MB deconvolution from (d), respectively (same as Figs. 4f and 4h, respectively).

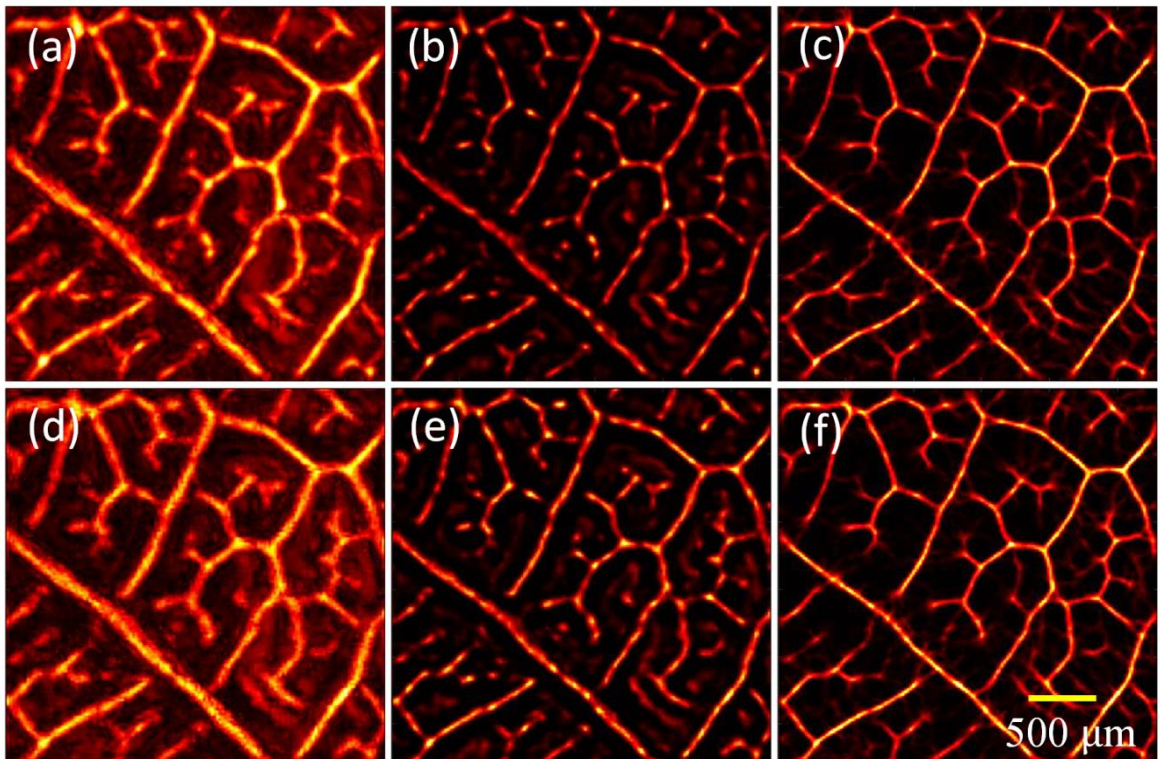

Fig. S5. Comparison of FA-SAFT-dir0 and FA-SAF-dir1 for their following deconvolution results. Leaf skeleton phantom is used. (a) By FA-SAFT-dir1 (same as Fig. 7f). (b,c) R-L deconvolution and D-MB deconvolution from (a), respectively. (d) By FA-SAFT-dir0 (same as Fig. 7g). (e,f) R-L deconvolution and D-MB deconvolution from (d), respectively (same as Figs. 7h and 7i, respectively).

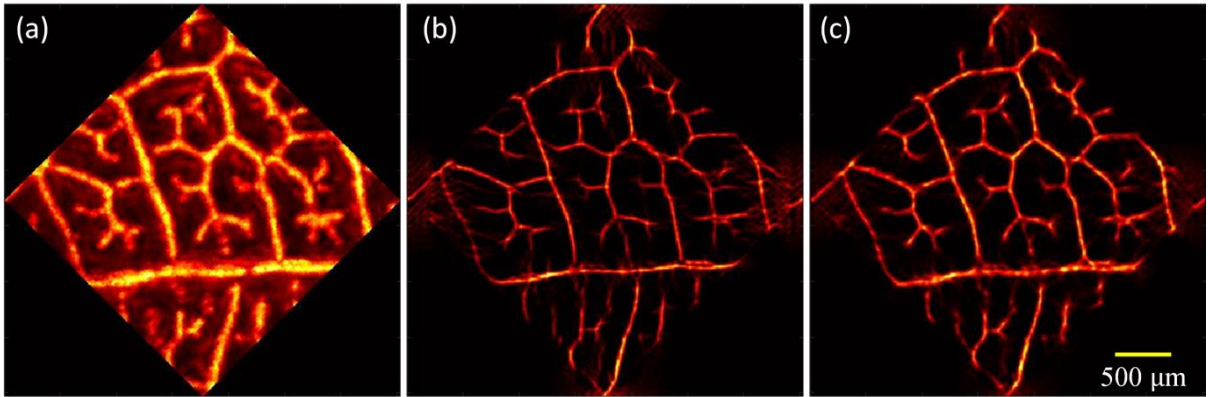

Fig. S6. (a) Fig. 7(a) rotated by 45°. (b) D-MB deconvolution (M = 1) from (a). (c) D-MB deconvolution (M = 4) from (a). Although (b) can also preserve most features, (c) shows better pattern continuity and SNR.

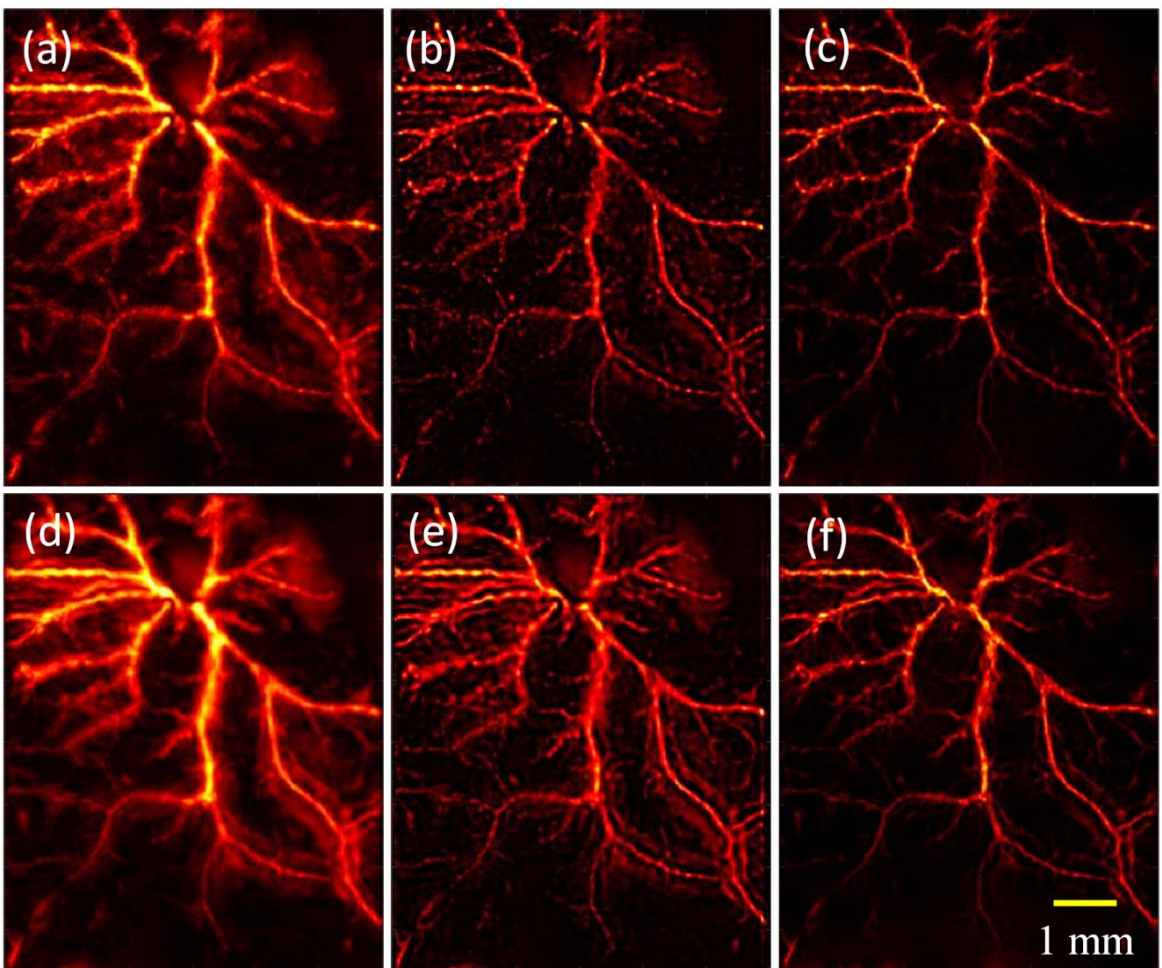

Fig. S7. Comparison of FA-SAFT-dir0 and FA-SAF-dir1 for their following deconvolution results. *In vivo* imaging of mouse dorsal blood vessels is used. (a) By FA-SAFT-dir1 (same as Fig. 8c). (b,c) R-L deconvolution and D-MB deconvolution from (a), respectively. (d) By FA-SAFT-dir0 (same as Fig. 8d). (e,f) R-L deconvolution and D-MB deconvolution from (d), respectively (same as Figs. 8e and 8f, respectively).

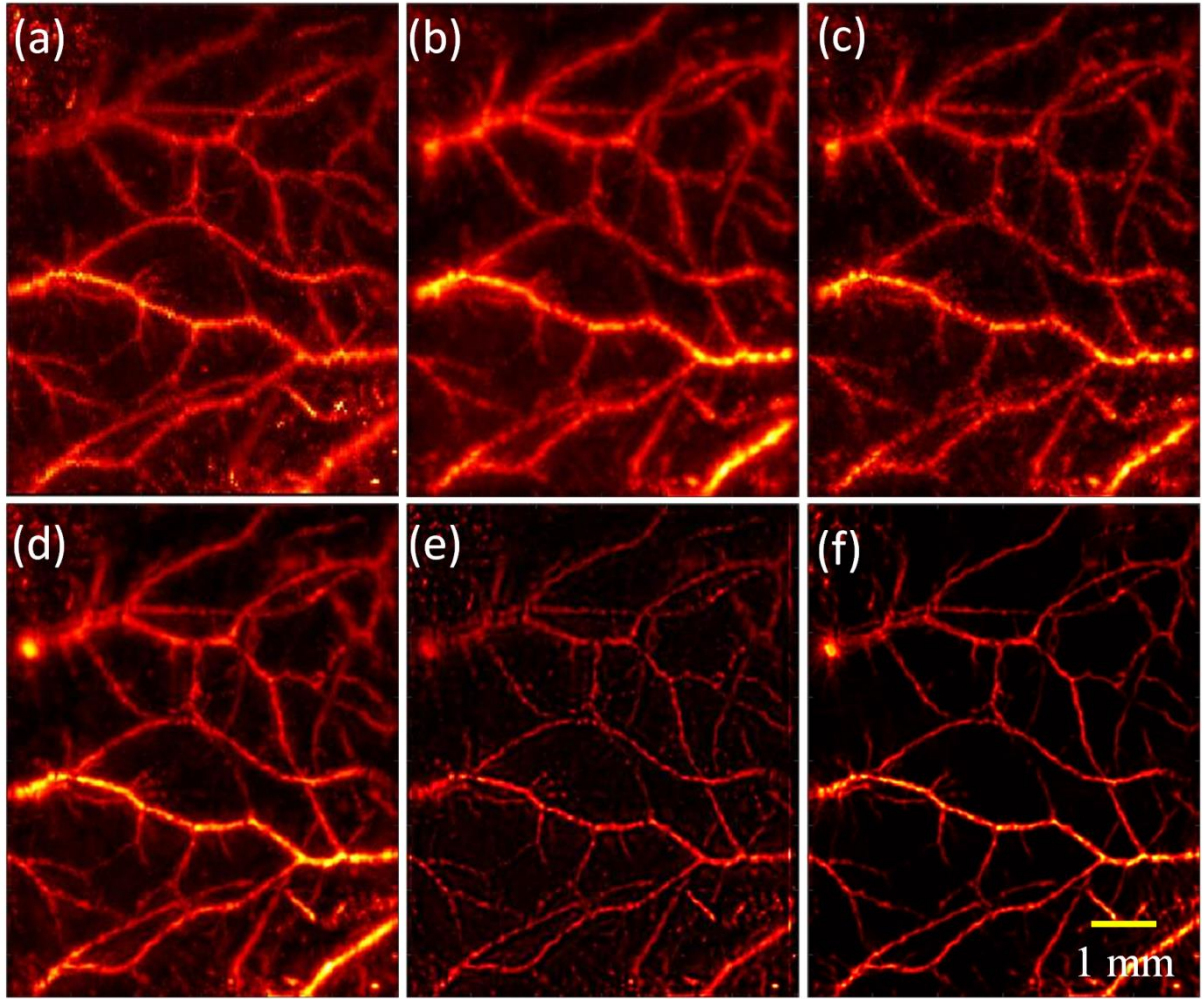

Fig. S8. Results of *in vivo* mouse experiment (another mouse dorsal blood vessels). (a) PAM image of raw data (out-of-focus). (b–d) By D-SAFT, FA-SAFT-dir1, and FA-SAFT-dir0, respectively. (e,f) R-L deconvolution and D-MB deconvolution from (d), respectively.

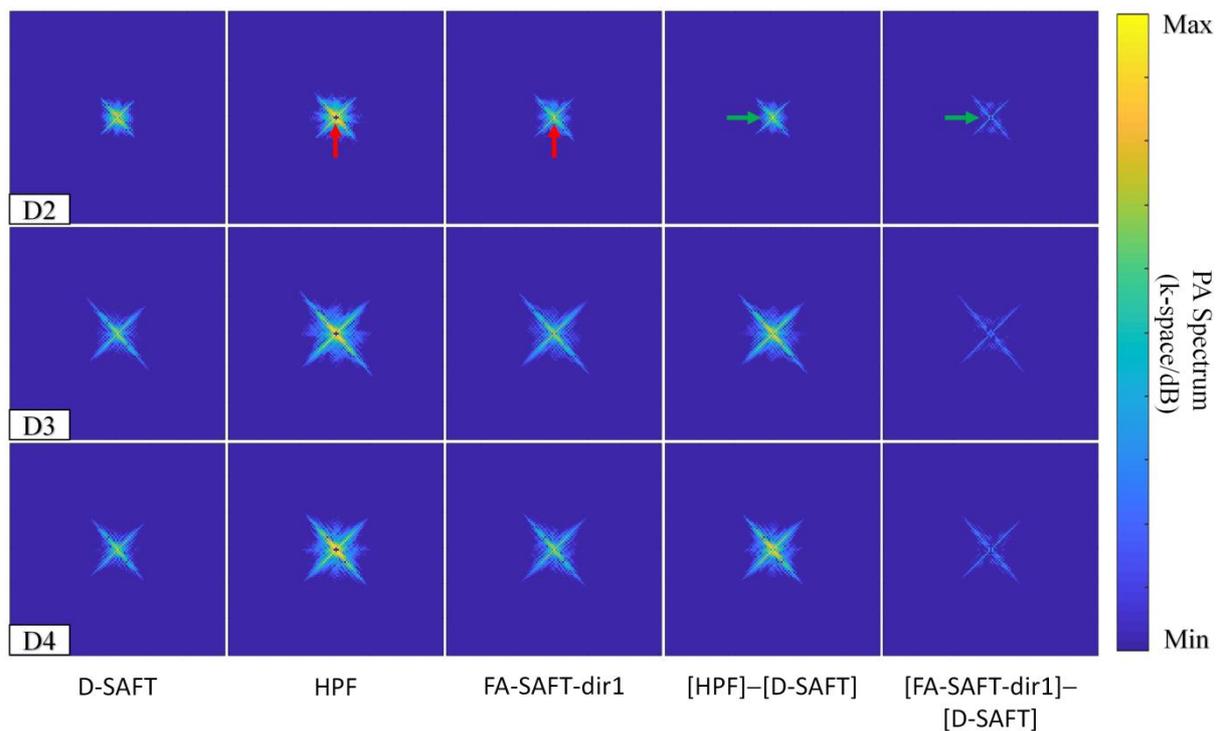

Fig. S9. Comparison of spectra among D-SAFT, HPF, and FA-SAFT-dir1. [HPF]−[D-SAFT], the spectrum division in linear scale (or difference in log scale (dB)) between HPF and D-SAFT. [FA-SAFT-dir1]−[D-SAFT], the spectrum division in linear scale (or difference in log scale (dB)) between FA-SAFT-dir1 and D-SAFT. Note that the spectra are plotted in log scale with dynamic range of 100 dB for better visualization.

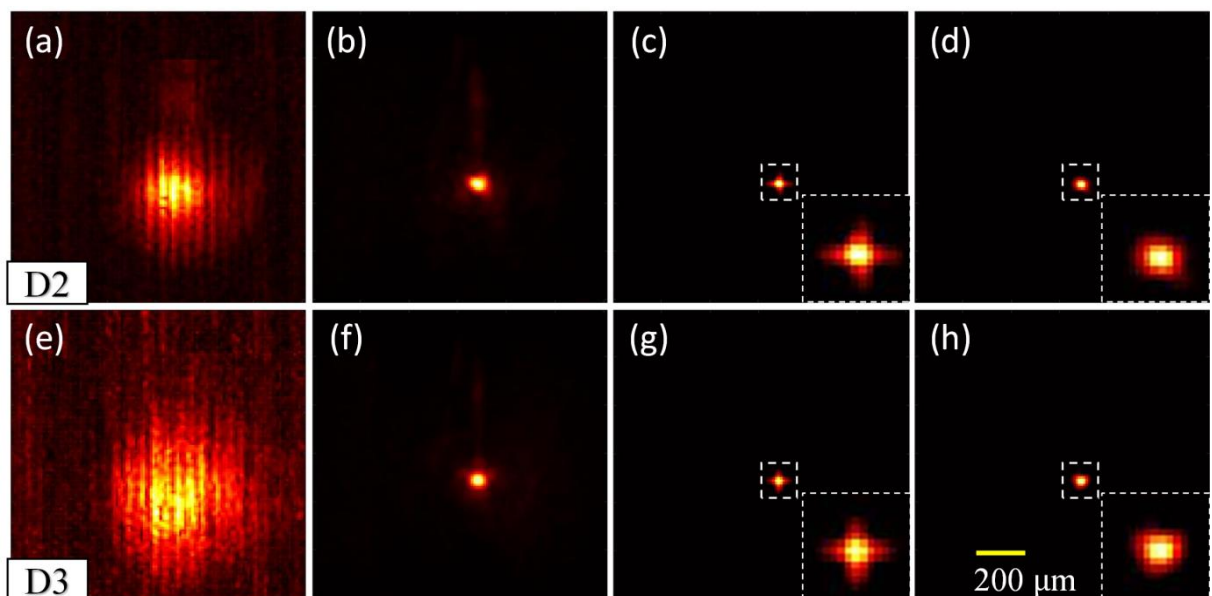

Fig. S10. Results of bead experiment. (a,e) MAP image of raw data at D2 (OFD of 0.3 mm) and D3 (OFD of 0.6 mm), respectively. (b,f) By FA-SAFT-dir1. (c,g) D-MB deconvolution (M = 1) from (b) and (f), respectively. (d,h) D-MB deconvolution (M = 4) from (b) and (f), respectively.

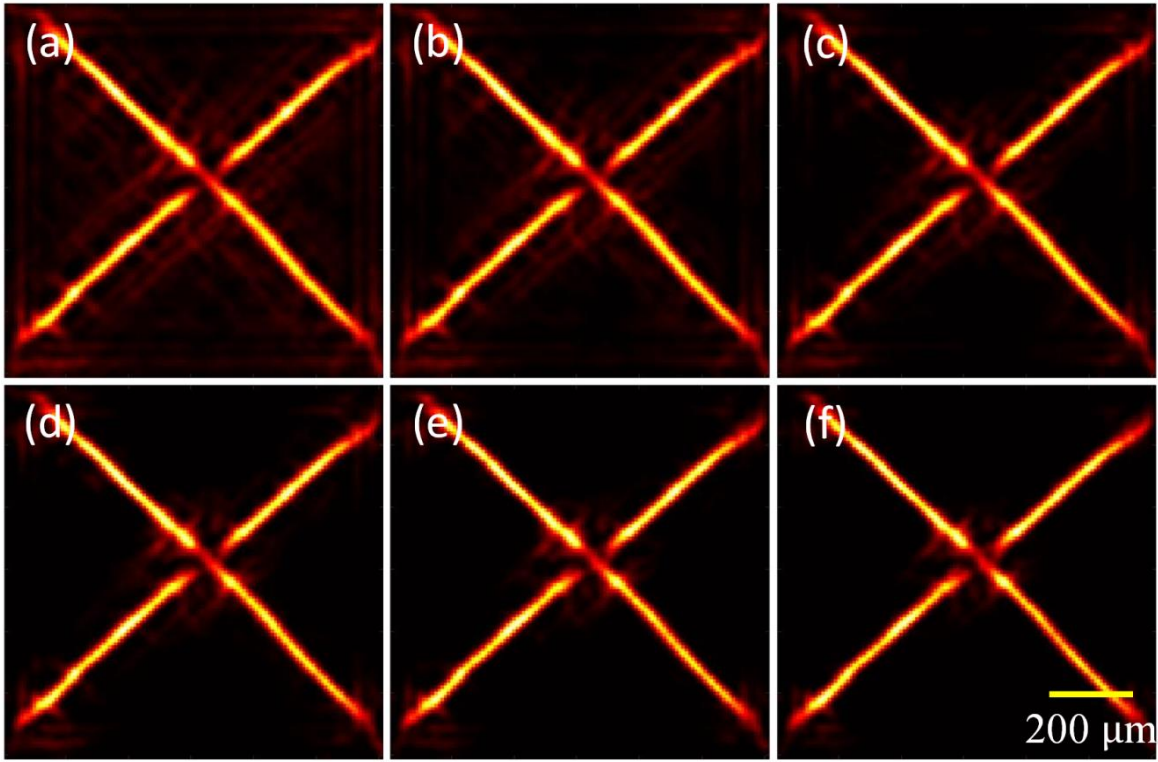

Fig. S11. Illustration of D-MB deconvolution results using different $\lambda$. The crossed tungsten wire phantom is used for illustration. (a–f) The results using $\lambda = 0$, 0.2, 0.4, 0.6, 0.8, and 1.0, respectively.

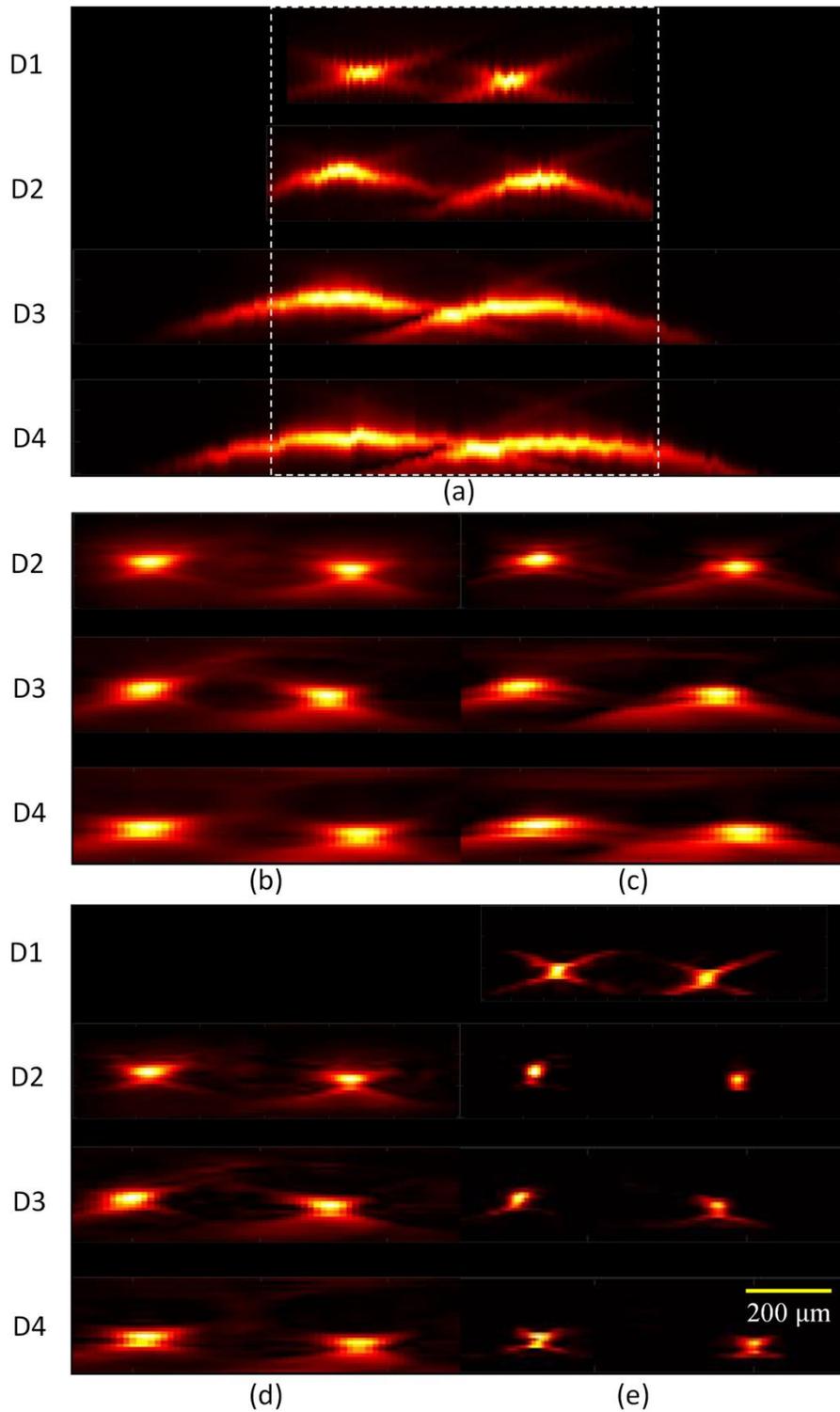

Fig. S12. Comparison of B-mode images of the crossed tungsten wires at different OFDs. (a) Raw data. The region in the dashed box is selected for the following comparisons in (b–e). (b) D-SAFT. (c) FA-SAFT-dir0. (d) FA-SAFT-dir1. (e) D-MB deconvolution. Lateral resolution is improved for objects at D2–D4 (i.e., DOF extended) by both D-SAFT and FA-SAFT (b–d) compared with raw data (a). Lateral resolution is further improved by D-MB deconvolution (e) compared with FA-SAFT-dir1 (d). Axial resolution comparison is shown in Table S1.

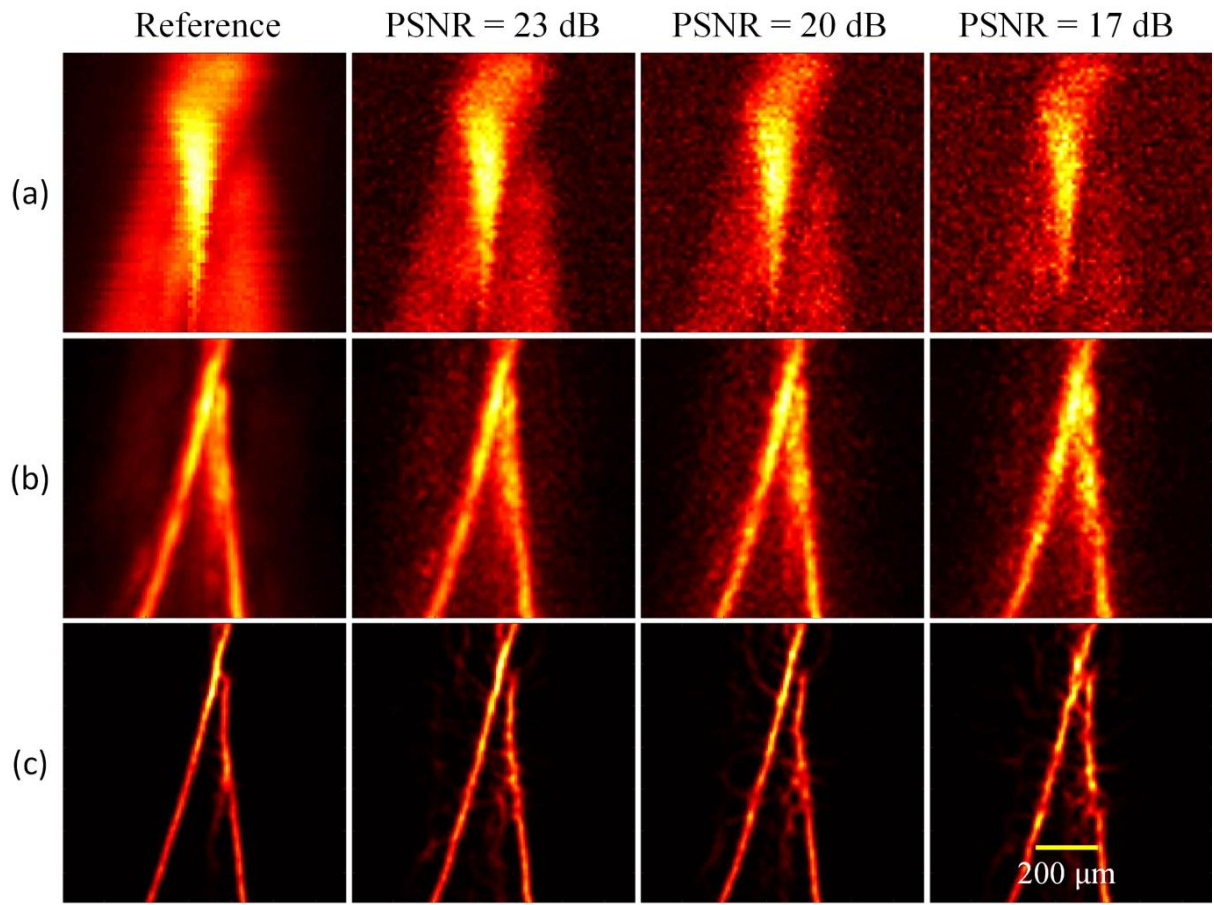

Fig. S13. The effect of SNR on the performance of FA-SAFT and D-MB deconvolution. Fig. 6d is chosen as a reference PA image with high PSNR (the top-left image). (a) Raw data with different PSNR (reference, 23 dB, 20 dB, and 17 dB from left to right, respectively), which are produced by adding different levels of white Gaussian noise to the PA A-line signals of the reference PA image. (b) FA-SAFT-dir1. (c) D-MB deconvolution from (b).

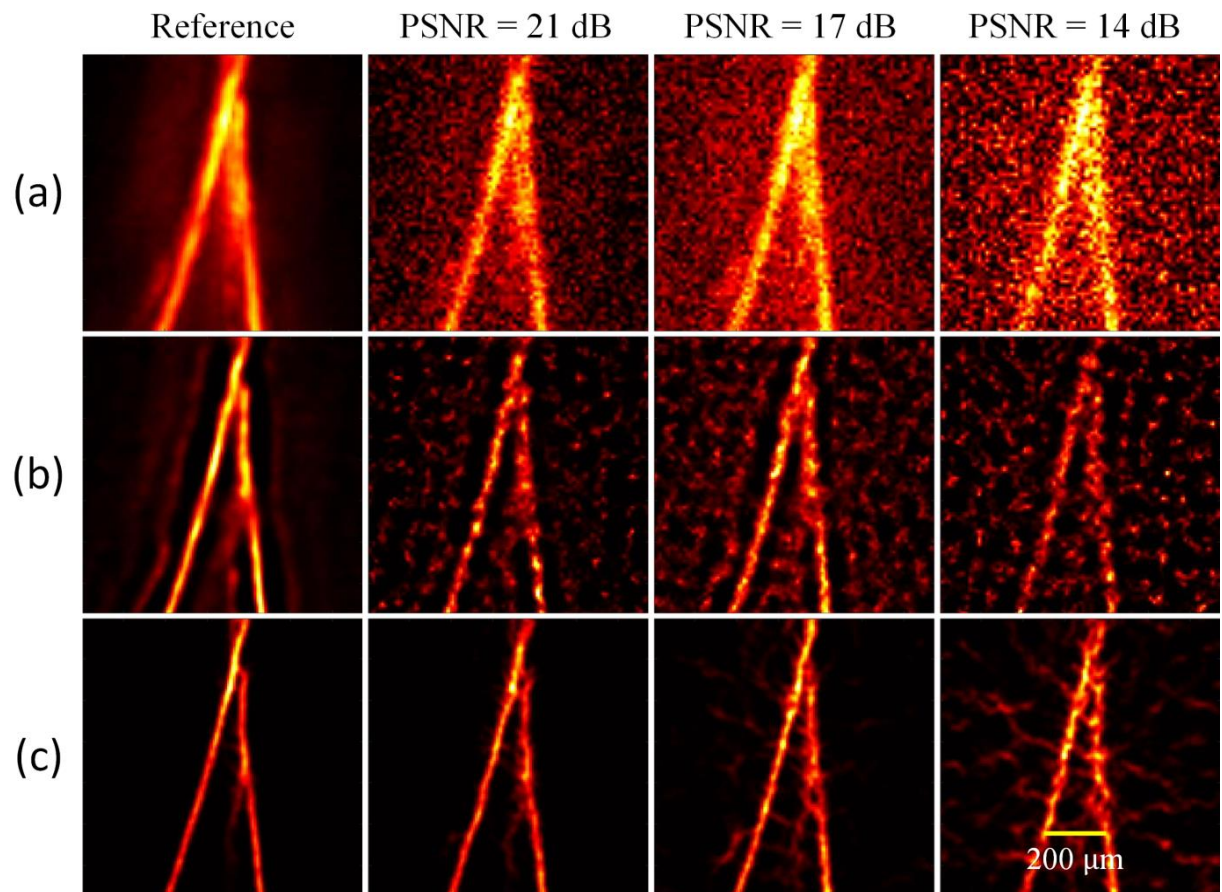

Fig. S14. The effect of SNR on the performance of D-MB deconvolution. Fig. 6e is chosen as a reference PA image with high PSNR (the top-left image). (a) FA-SAFT-dir1 with different PSNR (reference, 21 dB, 17 dB, and 14 dB from left to right, respectively), which are produced by adding different levels of white Gaussian noise to the 2D MAP of the reference PA image. (b) R-L deconvolution from (a). (c) D-MB deconvolution from (a).

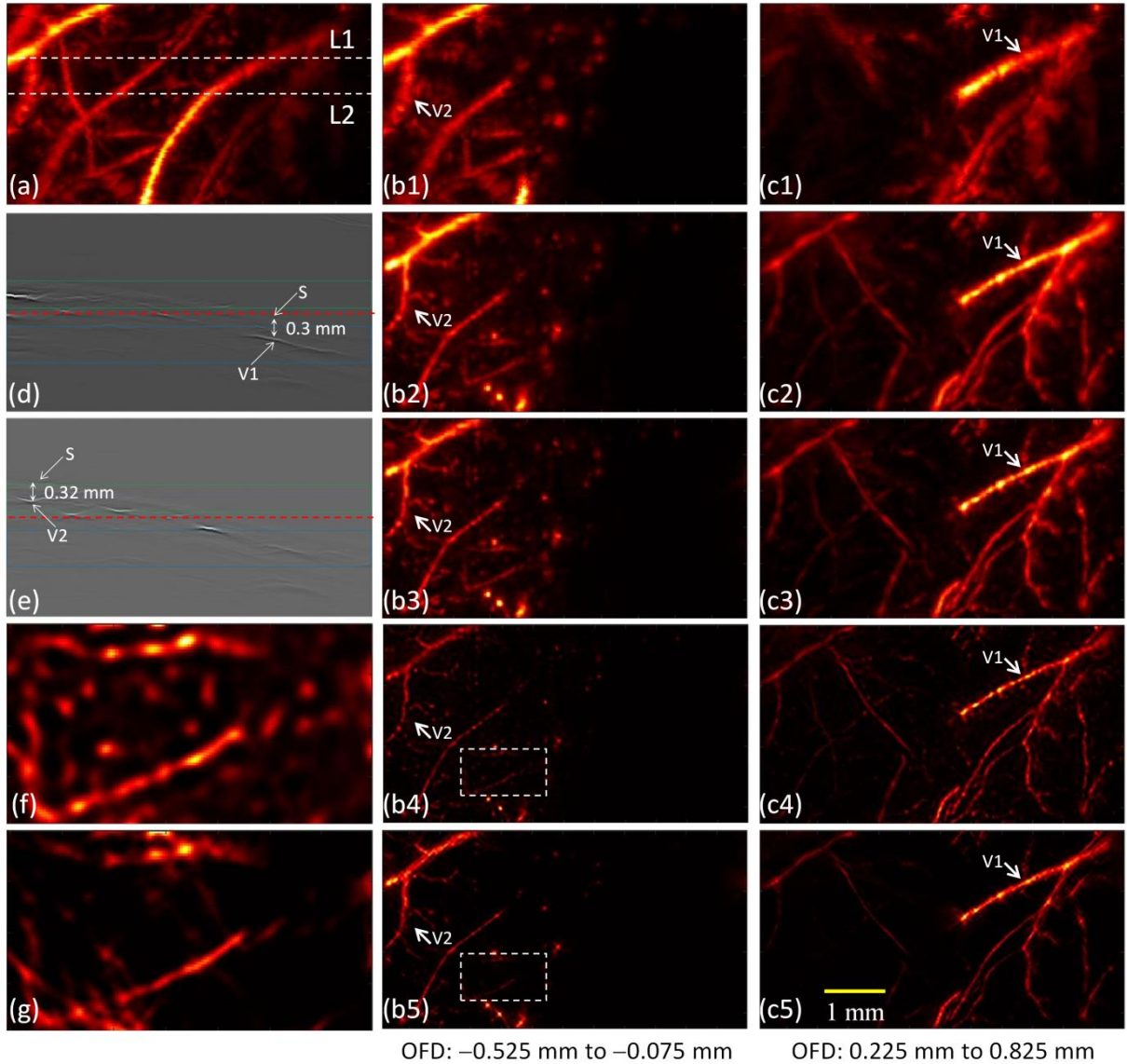

Fig. S15. The performance of the proposed algorithm when applied to images with penetration depth in tissue. A large axial range of 3.3 mm (OFD from −1.65 mm to 1.65 mm) is acquired. (a) PAM image of raw data (MAP over the whole axial range of 3.3 mm). (b1−b5) PAM images in MAP over the axial range of 0.45 mm above the focus (OFD from −0.525 mm to −0.075 mm). From top to bottom: (b1) PAM image of raw data, (b2,b3) by D-SAFT and FA-SAFT-dir1 from (b1), respectively, (b4,b5) by R-L deconvolution and D-MB deconvolution from (b3), respectively. (c1−c5) PAM images in MAP over the axial range of 0.6 mm below the focus (OFD of 0.225 mm to 0.825 mm). From top to bottom: (c1) PAM image of raw data, (c2,c3) by D-SAFT and FA-SAFT-dir1 from (c1), respectively, (c4,c5) by R-L deconvolution and D-MB deconvolution from (c3), respectively. (d,e) B-mode images of raw data along the lines L1 and L2 in (a), respectively. The red line indicates the focal plane of the transducer (i.e., OFD of 0 mm), the green box indicates the axial range of 0.45 mm above the focus for (b1−b5), and the blue box indicates the axial range of 0.6 mm below the focus for (c1−c5). In (d), the pointed vessel V1 is ~0.3 mm below the skin surface S, and the corresponding vessel in PAM images is indicated by the arrows in (c1−c5). In (e), the pointed vessel V2 is ~0.32 mm below the skin surface S, and the corresponding vessel in PAM images is indicated by the arrows in (b1−b5). As can be seen in (b1−b5) and (c1−c5) such as V1 and V2, the proposed algorithm demonstrates good performance for the *in vivo* AR-PAM images with penetration depth of ~0.32 mm. (f,g) Zoom-in images of (b4) and (b5) (dashed boxes), respectively, which is for better displaying the region with relatively low PA amplitudes.

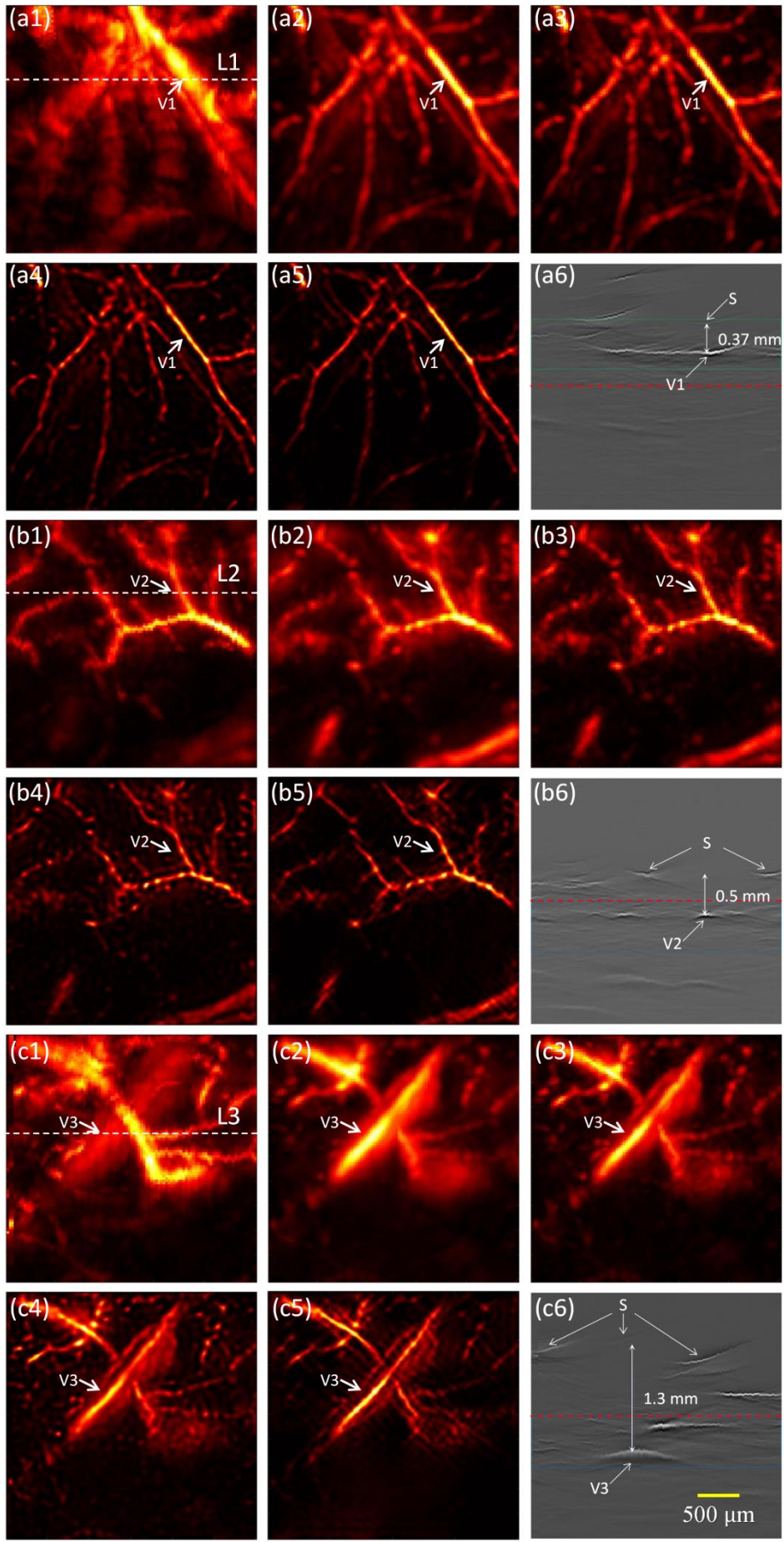

Fig. S16. The performance of the proposed algorithm when applied to images with penetration depth in tissue. Deeper blood vessels are intentionally sought in AR-PAM imaging experiments. (a1−a5) PAM images in MAP over the axial range of 0.6 mm above the focus (OFD from −0.8 mm to −0.2 mm). The images: (a1) PAM image of raw data, (a2,a3) by D-SAFT and FA-SAFT-dir1 from (a1), respectively, (a4,a5) by R-L deconvolution and D-MB deconvolution from (a3), respectively. (a6) B-mode image of raw data along the line L1 in (a1). The red line indicates the focal plane of the transducer (i.e., OFD of 0 mm) and the green box indicates the axial range of 0.6 mm above the focus for (a1−a5). In (a6), the pointed vessel V1 is ~0.37 mm below the skin surface S, and the corresponding vessel in PAM images is indicated by the arrows in (a1−a5). (b1−b5) PAM images in MAP over the axial range of 0.6 mm below the focus (OFD from 0 mm to 0.6 mm). The images: (b1) PAM image of raw data, (b2,b3) by D-SAFT and FA-SAFT-dir1 from (b1), respectively, (b4,b5) by R-L deconvolution and D-MB deconvolution from (b3), respectively. (b6) B-mode image of raw data along the line L2 in (b1). The red line indicates the focal plane of the transducer (i.e., OFD of 0 mm) and the blue box indicates the axial range of 0.6 mm below the focus for (b1−b5). In (b6), the pointed vessel V2 is ~0.5 mm below the skin surface S, and the corresponding vessel in PAM images is indicated by the arrows in (b1−b5). (c1−c5) PAM images in MAP over the axial range of 0.6 mm below the focus (OFD from 0 mm to 0.6 mm). The images: c1) PAM image of raw data, (c2,c3) by D-SAFT and FA-SAFT-dir1 from (c1), respectively, (c4,c5) by R-L deconvolution and D-MB deconvolution from (c3), respectively. (b6) B-mode image of raw data along the line L3 in (c1). The red line indicates the focal plane of the transducer (i.e., OFD of 0 mm) and the blue box indicates the axial range of 0.6 mm below the focus for (c1−c5). In (c6), the pointed vessel V3 is up to ~1.3 mm below the skin surface S, and the corresponding vessel in PAM images is indicated by the arrows in (c1−c5). As can be seen in the blood vessel V3 in (c1−c6), the proposed algorithm still demonstrates the effectiveness of resolution enhancement for the *in vivo* AR-PAM image of deep vessels up to ~1.3 mm below the skin surface. A few artifacts observed in (c4) and (c5) could be due to relatively poor image quality in (c1).